\documentclass[aps,pra,showpacs,twocolumn,superscriptaddress]{revtex4-1}

\usepackage{amsmath}
\usepackage{graphicx}
\usepackage{bm}

\newcommand{\ua}{\uparrow}
\newcommand{\da}{\downarrow}

\begin{document}

\title{Dynamical symmetry and pair tunneling in a one-dimensional Bose gas colliding with a mobile impurity}
\author{Elmer V. H. Doggen}
\email[Corresponding author: ]{elmer.doggen@aalto.fi}
\affiliation{COMP Centre of Excellence and Department of Applied Physics, Aalto University, FI-00076 Aalto, Finland}
\author{Sebastiano Peotta}
\affiliation{COMP Centre of Excellence and Department of Applied Physics, Aalto University, FI-00076 Aalto, Finland}
\author{P\"aivi T\"orm\"a}
\affiliation{COMP Centre of Excellence and Department of Applied Physics, Aalto University, FI-00076 Aalto, Finland}
\affiliation{Institute for Quantum Electronics, ETH Z\"urich, 8093 Z\"urich, Switzerland}
\author{Jami J. Kinnunen}
\affiliation{COMP Centre of Excellence and Department of Applied Physics, Aalto University, FI-00076 Aalto, Finland}

\begin{abstract}
Using the time-dependent density matrix renormalization group (TDMRG) we theoretically study the collision of a one-dimensional gas of interacting bosons with an impurity trapped in a shallow box potential.
We study the dynamic response of both the impurity and the bosonic gas and find an approximate independence of the sign of the interaction (attractive or repulsive) between the impurity and the bosons.
This sign-independence breaks down when the interaction energy is of the same order as the box potential depth, leading to resonant pair tunneling.
Our predictions can be tested using contemporary techniques with ultracold atoms.
\end{abstract}


\maketitle

\emph{Introduction.---}
Ultracold atoms are an ever more versatile toolbox for simulating complex condensed matter systems as well as for experimentally realizing various theoretical ``toy models.''
In particular, there has been significant interest in one-dimensional (1D) systems over the previous decade.
Such systems were previously regarded mainly as a mathematical curiosity, but were realized experimentally in the early 2000s \cite{Moritz2003a,Kinoshita2004a,Paredes2004a}.
One-dimensional systems can be created by tightly confining a dilute cloud of atoms in two of the three dimensions \cite{Olshanii1998a} and the strength of the inter-particle interactions can be tuned using Feshbach resonances \cite{Kokkelmans2014a} and confinement-induced resonances \cite{Bergeman2003a,Haller2010a}.
Subsequent experimental efforts in 1D include various realizations of Bose gases and the dynamics therein \cite{vanAmerongen2008a,Haller2009a,Fabbri2015a,Kinoshita2006a,Hofferberth2007a,Gring2012a,Trotzky2012a,Ronzheimer2013a,Langen2013a}, studies of the dynamics of impurities \cite{Palzer2009a,Spethmann2012a,Catani2012a,Fukuhara2013a}, and fermionic few-body systems \cite{Zurn2012a, Wenz2013a, Zurn2013a}.
Theoretically, one-dimensional systems are of interest because of the existence of exact solutions and powerful approximative techniques for various relevant many-body scenarios -- see, e.g.,  Refs.\ \cite{Giamarchi2003a,Cazalilla2011a,Imambekov2012a,Guan2013a} for reviews.
In addition, the ground state solutions of several classes of one-dimensional Hamiltonians and the dynamics of particles described thereby can be calculated numerically essentially to arbitrary accuracy using the density matrix renormalization group \cite{Schollwock2011a}, time-evolving block decimation \cite{Vidal2003a} or exact diagonalization of the Hamiltonian \cite{Zhang2010a}.

The effect of an impurity on the properties of a 1D quantum gas has been extensively studied \cite{Johnson2012a,Knap2012a,Schecter2012a,Mathy2012a,Bonart2012a,Massel2013a,Sindona2013a,Peotta2013a,Kantian2014a,Burovski2014a,Visuri2014a}; in higher dimensions one can experimentally realize the so-called \emph{Fermi polaron} \cite{Nascimbene2009a, Schirotzek2009a}, described accurately using a simple variational model \cite{Chevy2006a}.
This quasiparticle description works even in 1D \cite{Giraud2009a,Doggen2013a,Doggen2014a}, even though strictly speaking these quasiparticles are not well-defined in 1D.
A bosonic analog, the \emph{Bose polaron}, has also been proposed \cite{Rath2013a,Dutta2013a,Li2014a} and experimentally realized in 1D \cite{Fukuhara2013a}.

\begin{figure}
 \vspace{-5mm}
 \includegraphics[width=\columnwidth]{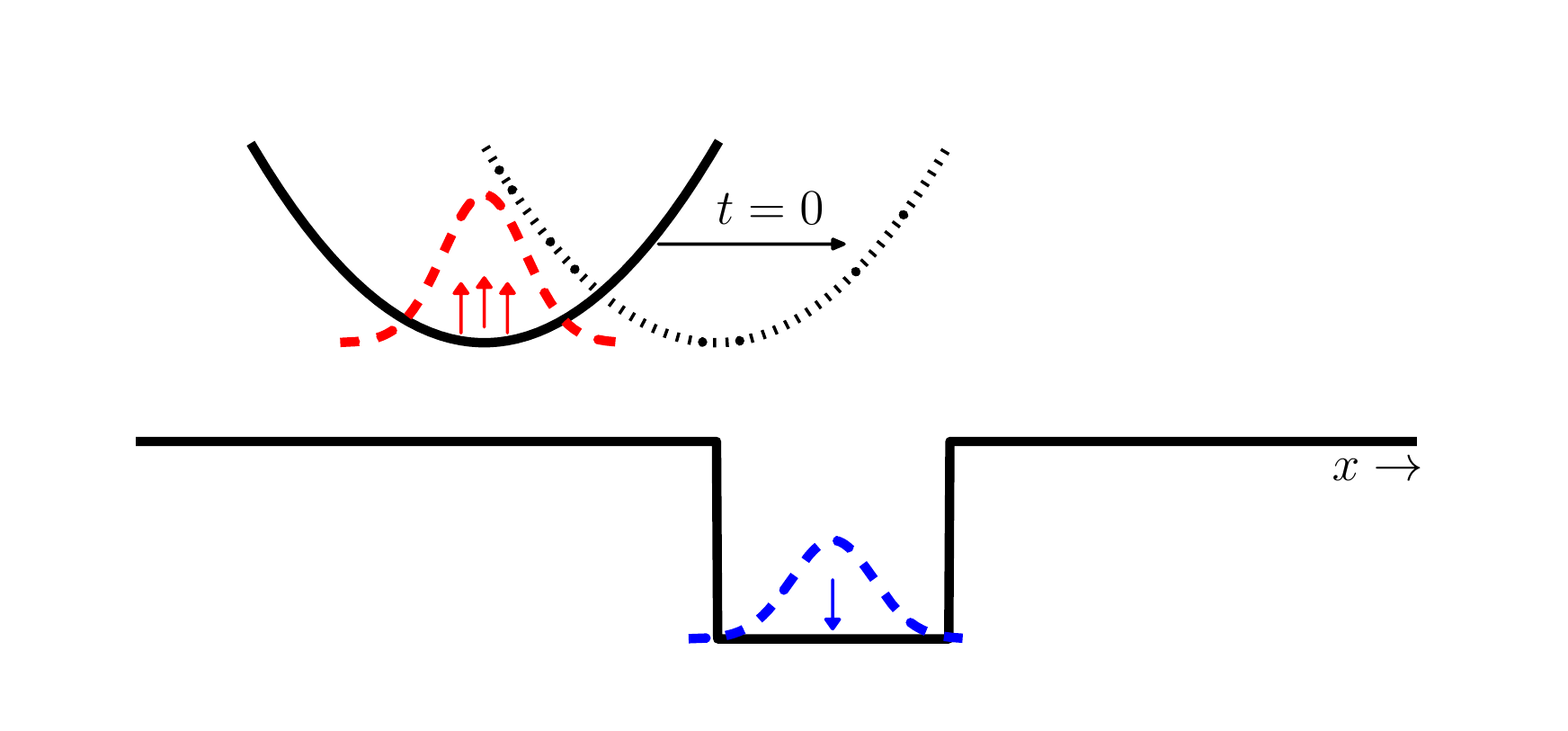}
 \vspace{-10mm}
 \caption{(color online). Schematic depiction of the problem under consideration.
 Initially, $N_\uparrow$ majority component bosons (red dashed line and arrows) are trapped in a harmonic trap, while a single $\downarrow$ impurity (blue dashed line and arrow) is trapped in a separate box potential.
 At a time $t = 0$, the harmonic trap is instantaneously displaced (dotted line), allowing the $\uparrow$-particles to collide and interact with the trapped impurity.
 $x$ indicates the one-dimensional position.}
 \label{fig:schematic}
\end{figure}

Here, we will consider a scenario depicted schematically in FIG.\ \ref{fig:schematic}.
An ensemble of interacting 1D bosons is initially trapped in a harmonic potential (for related theoretical studies of 1D bosons see Refs.\ \cite{Li2003a,Fuchs2005a,Guan2007a,Tempfli2008a,Tempfli2009a,Caux2009a,Brouzos2012a,Vidmar2013a,Campbell2014a,Boschi2014a,Zinner2014a,Campbell2015a}).
At a time $t=0$, the harmonic trap is displaced, allowing the bosons to collide and interact with an impurity.
This entails a quantum \emph{quench} (a sudden change in one of the parameters of the Hamiltonian); such quenches have been studied in particular in the context of bridging the gap between the micro-world of quantum physics and macroscopic statistical ensembles \cite{Rigol2007a,Rigol2008a,Cazalilla2010b,Polkovnikov2011a}.
We place the impurity in a finite well or ``box'' potential \cite{Meyrath2005a} and investigate the tunneling dynamics \cite{Zollner2008a,Rontani2012a,Zurn2013a,Simpson2014a,Lundmark2015a} and spin transport \cite{Sommer2011a} of the particle outside the box as a function of the various parameters of the setup, such as the interaction between the majority bosons and the depth of the box potential.
Recently, a similar system consisting of a few bosons interacting with an impurity on the sites of an optical lattice was studied experimentally \cite{Will2011a}.

\begin{figure*}[!htb]
 \includegraphics[width=18cm]{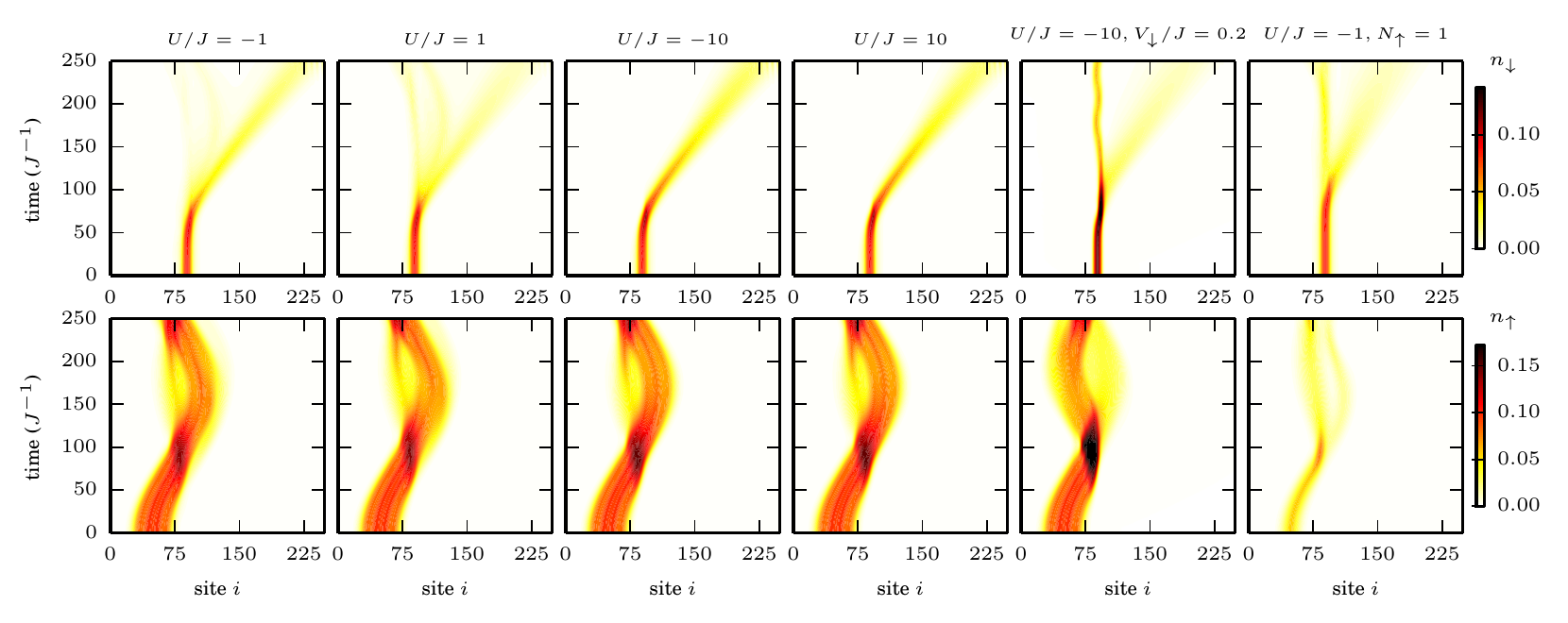}
 \caption{(color online). Density profiles for the impurity $n_\da$ (top panels) and the majority component bosons $n_\ua$ (lower panels) and interaction strengths (columns) $U/J = \pm 1, \pm 10$. In the panels, the $x$-axis indicates position and the $y$-axis time. Darker colors indicate higher densities. Note the many-body effects for $U/J = \pm 1$ as evidenced by the formation of ``branches'' in the density profile. In the case of a single majority component particle (right panels), only two such branches are visible. A ``quantum Newton's cradle''-type effect is visible in the majority component response \cite{Kinoshita2006a} (lower panels, first four columns). For strong interactions $U/J = \pm 10$, the impurity is fully kicked out of the box potential, unless it is sufficiently strongly trapped (fifth column). The number of majority component bosons $N_\ua = 4$, the intra-species interaction $U_\ua/J = 5$ and the depth of the box potential $V_\downarrow/J = 0.05$ unless otherwise noted. See also the online supplementary material for animated density profiles.}
 \label{fig:densities}
\end{figure*}

\emph{Model.---}
We consider a one-dimensional system at zero temperature with an impurity ($\da$) interacting with a number of ($\ua$) bosons, which themselves can interact with each other.
This system can be described using the Bose-Hubbard Hamiltonian for a two-component system:
\begin{align}
 \hat{H} = & -J \sum_{\sigma i} b_{\sigma i}^\dagger b_{\sigma i+1} + \text{H.c.} + U_\ua \sum_i \hat{n}_{\ua i} (\hat{n}_{\ua i} - 1) \nonumber \\
 & + U \sum_i \hat{n}_{\ua i} \hat{n}_{\da i} + \sum_{\sigma i} V_{\sigma i} \hat{n}_{\sigma i} \label{eq:BoseHub}
\end{align}
Here $J$ represents the hopping parameter, $U_\ua$ is the on-site interaction energy between the majority component $\ua$-bosons, $U$ represents the interaction between the impurity and the $\ua$-bosons and $\hat{n}_{\sigma i} = b_{\sigma i}^\dagger b_{\sigma i}$ is the density operator, where $b_{\sigma i}^{(\dagger)}$ is the bosonic annihilation (creation) operator for a species of generalized spin $\sigma \in \{\ua,\da\}$ at lattice site $i$.
The external potential is added through an on-site term $V_{\sigma i}$, which is different for both species.
We consider only the case where the on-site density $\langle \hat{n}_\sigma \rangle_i \ll 1 \, \forall i$, in which case the Bose-Hubbard Hamiltonian can be mapped to its continuum equivalent, the (two-component) Lieb-Liniger gas with contact ($\delta$-function) interactions \cite{Lieb1963a,Cazalilla2011a} with added terms for the external potentials.
In this mapping, the hopping parameter $J$ is mapped to the mass $m$ of a particle, which we assume to be the same for both species.
The on-site interaction energies $U_\ua$ and $U$ can be associated with the strength of the intra- and inter-species contact interactions respectively.

\emph{Results.---}
We use the time-dependent density matrix renormalization group ({\small TDMRG}) \cite{Schollwock2011a}.
This method yields a numerically exact solution for the ground state and time dynamics of the Bose-Hubbard Hamiltonian \eqref{eq:BoseHub}.
Although we are interested primarily in the continuum limit, we will state the Hubbard model parameters for transparency.
Our quench protocol is depicted schematically in FIG.\ \ref{fig:schematic}.
We use a system of $L = 250$ sites, intra-species interactions $U_\ua/J = 5$ and vary the inter-species interaction $U$ and number of majority component bosons $N_\ua$.
The choice $U_\ua/J = 5$ approximately corresponds to a Lieb-Liniger parameter $\gamma_{\text{LL}} = U_\ua/2J\tilde{n}_\ua \approx 25$, using half the peak density $\tilde{n}_\ua$ of the majority bosons.
Initially, the majority bosons are trapped in a harmonic trap $V_{\ua i}/J = 10^{-4}(x_\text{A}-i)^2$ and the impurity is trapped in a box potential of depth $V_\da/J = 0.05$ for $85 \leq i \leq 95$.
We prepare these subsystems in their respective ground states with $U = 0$.
To elucidate the effects of the interactions between the particles, we assume that only the impurity ``feels'' the trapping potential of the box potential, while only the majority bosons feel the harmonic trap.
Such a setup might be realized experimentally using species-selective trapping potentials \cite{LeBlanc2007a}.
The harmonic trap is displaced from $x_\text{A} = 50$ to $x_\text{B} = 80$ at a time $t=0$, while at the same time the inter-species interaction $U$ is switched on.
This choice ensures that the initial overlap between the two clouds is small enough so that the effect of the \emph{interaction} quench is not relevant.
Furthermore, this particular quench protocol implies that, in the continuum limit and in the absence of interactions with the impurity, the cloud of majority component particles will perform harmonic motion about $x_\text{B}$ indefinitely.
This allows us to see precisely the effect of the impurity on the majority component particles.
The impurity is initially only weakly confined; the box potential permits exactly one single-particle bound state for $V_\da/J = 0.05$.
After the displacement of the harmonic trap, we follow the dynamics of the system until $t = 250 J^{-1}$.
This time corresponds approximately to the period of the oscillation of the majority component bosons in the harmonic trap.
The density profiles for the impurity $n_\da \equiv \langle \hat{n}_{\da} \rangle_i$ and the majority bosons $n_\ua \equiv \langle \hat{n}_{\ua} \rangle_i$ as a function of time for various parameters are depicted in FIG.\ \ref{fig:densities}.

\begin{figure}[!htb]
 \includegraphics[width=\columnwidth]{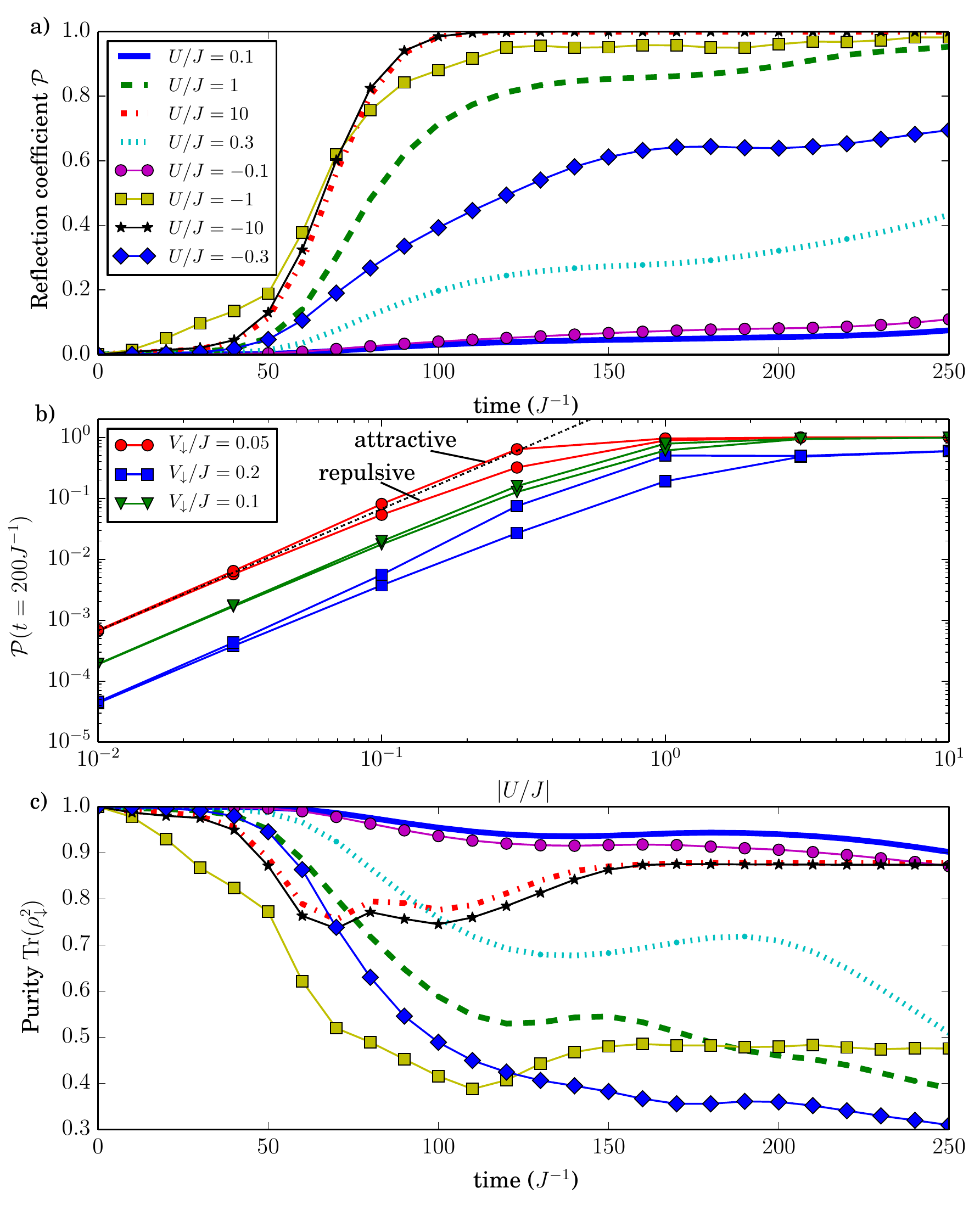}
 \caption{(color online). \textbf{a)} probability $\mathcal{P}$ that the impurity is kicked outside the box potential, eq.\ \eqref{eq:prob}. \textbf{b)} (note the double log scale) $\mathcal{P}$ evaluated at a fixed time $t = 200 J^{-1}$ for various interaction strengths $U/J$ and box potential depths $V_\da/J$. Both positive (lower lines) and negative (upper lines) values of $U/J$ are shown. The dotted line is a guide to the eye and is proportional to $U^2$. Note that for $V_\da/J = 0.1, 0.2$ we subtracted the overlap with the first \emph{two} single-particle eigenstates. \textbf{c)} purity $\gamma$ of the reduced density matrix $\rho_\da$, Eq.\ \eqref{eq:purity}, as a function of time. The strongest entanglement of the impurity and the majority component bosons occurs for intermediate interactions $U/J = \pm 0.3,1$. $N_\ua = 4, V_\da/J = 0.05$ for all panels unless noted otherwise.}
 \label{fig:overlap}
\end{figure}

After the displacement of the harmonic trap, the cloud of the majority propagates towards the impurity, which then has a certain probability of being kicked out of the shallow box trap.
We can quantify this probability by computing the following quantity:
\begin{equation}
 \mathcal{P} = 1 - \sum_{ij} \phi_i \rho_{\da ij} \phi_j^*, \label{eq:prob}
\end{equation}
where $\rho_\da$ is the reduced density matrix of the $\da$-particle and $\phi$ is the single-particle ground state of the impurity obtained through an exact diagonalization method independent from our {\small TDMRG} implementation.
This quantity is equal to $0$ initially as the impurity is found in the ground state of the box potential, and reaches values of up to unity in the case the particle is completely kicked outside the box potential. 
In addition, we track the time evolution of the purity of the reduced density matrix:
\begin{equation}
 \gamma = \text{Tr}( \rho_\da^2 ). \label{eq:purity}
\end{equation}
This quantity enables us to quantify the degree of entanglement between the majority bosons and the impurity.
It is equal to unity at $t = 0$ since the system is prepared such that the impurity and the $\ua$-bosons are independent, i.e.\ the initial state is a pure state.
The quantities \eqref{eq:prob} and \eqref{eq:purity} are depicted in FIG.\ \ref{fig:overlap}.

Interestingly, the behavior of the probability $\mathcal{P}$ can be described remarkably well using a simple two-particle picture.
In free space, the reflection coefficient (the square of the scattering amplitude) $|f(k)|^2$ for a particle with momentum $k$ interacting through a contact potential $u(|x-x'|) = g\delta(|x-x'|)$ with a second, stationary particle at position $x'$ is given by \cite{Olshanii1998a,Goulko2011a,Ozaki2012a} (we choose units where $\hbar = m/2 = 1$):
\begin{equation}
 |f(k)|^2 = \frac{g^2}{g^2 + k^2}. \label{eq:scattering}
\end{equation}
We should therefore expect the tunneling of the impurity outside of the trap, i.e.\ the reflection of the impurity from the majority component particles, to scale as $U^2$ for weak interactions (as might also be expected from a perturbative approach), with a saturation to complete reflection for $|U/J| \rightarrow \infty$.
This behavior is recovered by the numerical results in the limits of weak and strong interaction.
Interestingly, an \emph{interaction} quench leads to similar results \cite{Doggen2014b}.
Note that this picture ignores the depth of the box potential confining the impurity; indeed, for a sufficiently deep potential (see FIG.\ \ref{fig:densities}, fifth column) tunneling of the impurity is suppressed, as expected.
In this case there is also an additional single-particle bound state in the box potential.
The effect of $V_\da$ is examined more closely in FIG.\ \ref{fig:overlap}b.
Note that the effect of the sign of the interaction is more pronounced for $V_\da/J = 0.2$.
In the weakly interacting limit, the $U^2$ scaling is recovered, albeit with a suppressed tunneling probability.
Meanwhile, in the strongly interacting limit $U \rightarrow \infty$, the tunneling probability saturates to a finite value smaller than unity.
However, the sign independence remains in both the weakly and strongly interacting limits.

For strongly attractive interactions, no significant pairing occurs between the $\ua$ bosons and the impurity, and the $\da$ particle is simply kicked outside the well with $\mathcal{P}$ approaching unity (corresponding to reflection), even though the static ground state of the system consists of the impurity pairing with one of the majority component bosons in the harmonic trap.
This effect was observed in an experiment with spin-polarized fermionic clouds \cite{Sommer2011a}.
A similar picture was used to analyze the collision of fermion clouds in Ref.\ \cite{Ozaki2012a}, although the sign of the interaction was not varied in this study.
In Ref.\ \cite{Goulko2011a}, a semiclassical Boltzmann approach was used to analyze the collision of two spin-polarized fermionic clouds, also predicting an independence of the sign of the interaction, while the authors of Ref.\ \cite{Taylor2011a} use a hydrodynamic approach and draw similar conclusions.
In lattice models, theoretical approaches \cite{Kajala2011a,Kajala2011b} and an experiment \cite{Schneider2012a} yield similar effects.
Note, however, that in the experiment of Ref.\ \cite{Schneider2012a} and the associated theory, the interaction sign independence relies on a specific symmetry of the lattice dispersion relation that is not present in the situation we discuss here.

In our simulations with spin-polarized bosons, we find the sign symmetry to hold over a wide range of interactions $U/J$.
However, the independence of the sign does not hold universally.
The most significant deviation occurs for intermediate interactions $U/J \approx 1$ \cite{Kajala2011a}.
This also manifests itself in the stronger entanglement of the impurity with the majority component bosons as shown in FIG.\ \ref{fig:overlap}c, as measured through the purity \eqref{eq:purity}.
We attribute this behavior to the relative importance of pair formation when the interaction energy is of the same order as the single-particle ground state energy $E_0$ in the box potential.
To quantify this, consider the mapping of the Bose-Hubbard Hamiltonian \eqref{eq:BoseHub} to the Lieb-Liniger Hamiltonian.
In this continuum mapping, we can associate the hopping $J$ with the mass of a particle through $J \rightarrow \hbar^2/2md$ \cite{Peotta2013a}, where $d$ is the lattice spacing, while $U \rightarrow g/d$, where $g$ measures the strength of the 1D contact interaction.
The energy of a bound state dimer of two particles interacting through a delta function potential is $-mg^2/4\hbar^2 = -U^2/8J$.
Equating this energy to $E_0$, which we can obtain numerically, we obtain for $V_\da/J = 0.05$ that $E_0/J \approx -0.027$, which yields $U/J \approx -0.46$, in good agreement with FIG.\ \ref{fig:overlap}b.
Meanwhile, for $V_\da/J = 0.2$ we obtain $U/J \approx -1.1$, confirming the shift to higher interaction strengths for deeper box potentials.
Thus, the independent collision model (ICM) of Ref.\ \cite{Ozaki2012a}, employing a series of independent, consecutive two-body collisions, breaks down when pairing effects are important.
For strong interactions $|U/J| \gg 1$ the purity \eqref{eq:purity} saturates at a finite value close to unity.
We explain this behavior through the impurity being kicked out of the box potential completely and then propagating as a free particle for later times.
Hence, while the interaction between the majority bosons and the impurity is strong, the induced entanglement is not.

\begin{figure}[!htb]
 \includegraphics[width=\columnwidth]{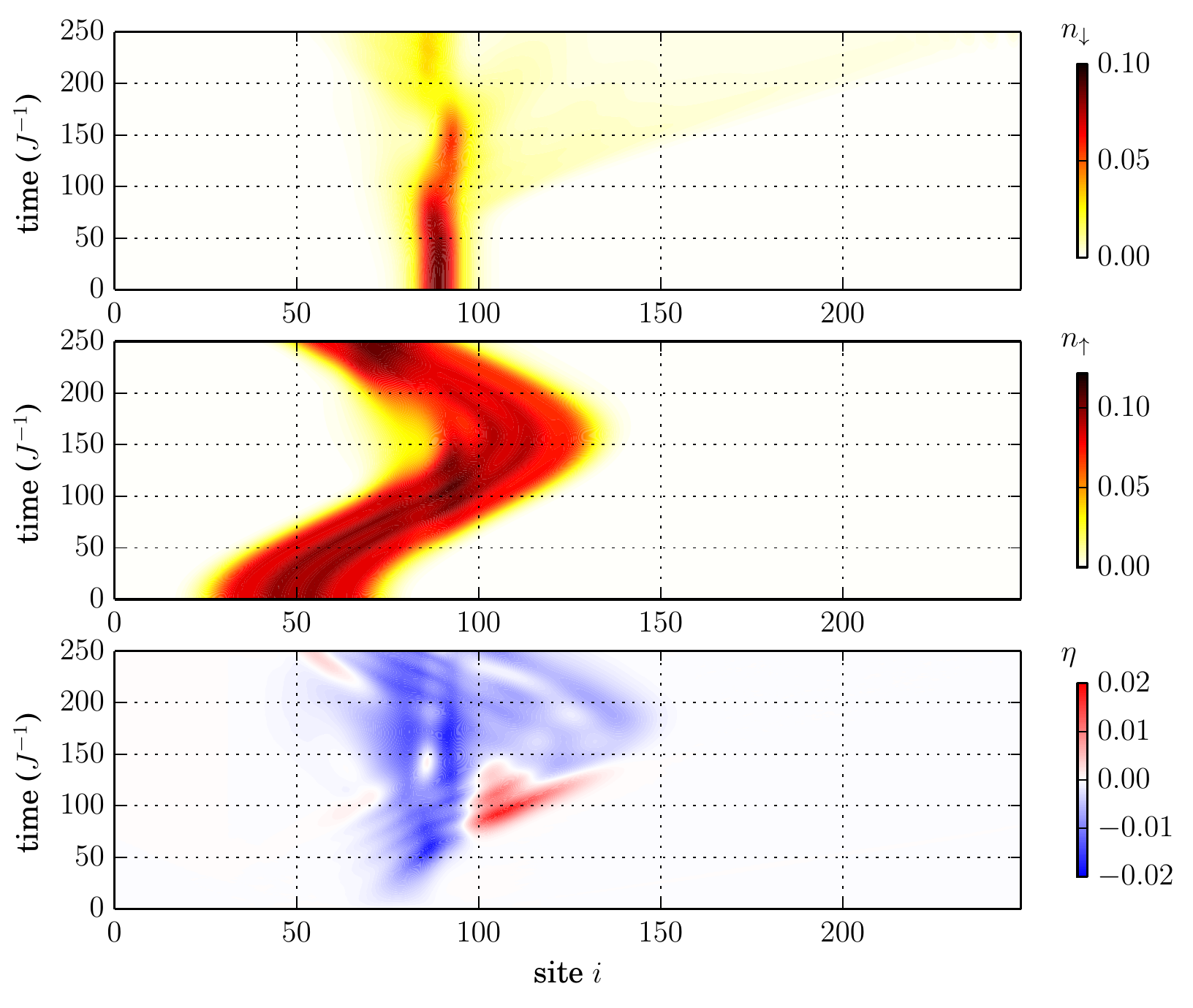}
 \caption{(color online). Density of the impurity $n_\da$ (top panel), the majority component bosons $n_\ua$ (middle panel) and the pair density $\eta$ (lower panel) as a function of time and position. Negative values in the lower panel indicate pairing, while positive values indicate anti-bunching. $U/J = -0.3$, $U_\ua/J = 5$, $V_\da/J = 0.05$ and $N_\ua = 4$.}
 \label{fig:doublonatt}
\end{figure}

The formed pairs are short-lived and dissociate due to the harmonic trapping that affects only the majority component bosons.
Therefore, the bosons act as an intermediary, enhancing the tunneling of the impurity out of the trap as is visible in the first two panels of the top row in FIG.\ \ref{fig:densities}.
We can visualize this effect by considering the pair density \cite{Kajala2011b}:
\begin{equation}
 \eta = \sqrt{n_\uparrow n_\downarrow} - n_{\downarrow \uparrow},
\end{equation}
where the square root of the doublon density $n_{\da \ua} \equiv \sqrt{\langle b_\da^\dagger b_\da b_\ua^\dagger b_\ua \rangle}$, and $n_\sigma = \langle b_\sigma^\dagger b_\sigma \rangle$ as before.
When $\eta < 0$, the density of doublons is greater than the product of densities, which indicates pairing.
Conversely, when $\eta > 0$ we have anti-bunching.
We consider the case where we expect the largest difference between repulsive and attractive interactions based on FIG.\ \ref{fig:overlap}b, i.e.\ $U/J = \pm 0.3$.
In FIG.\ \ref{fig:doublonatt} (lower panel) the process of pairing followed by dissociation is clearly visible.
Initially, when the bosonic cloud collides with the impurity, pairs are formed as indicated by the negative $\eta$.
The pair propagates briefly before dissociating, which coincides with positive $\eta$ at later times.

\begin{figure}[!htb]
 \includegraphics[width=\columnwidth]{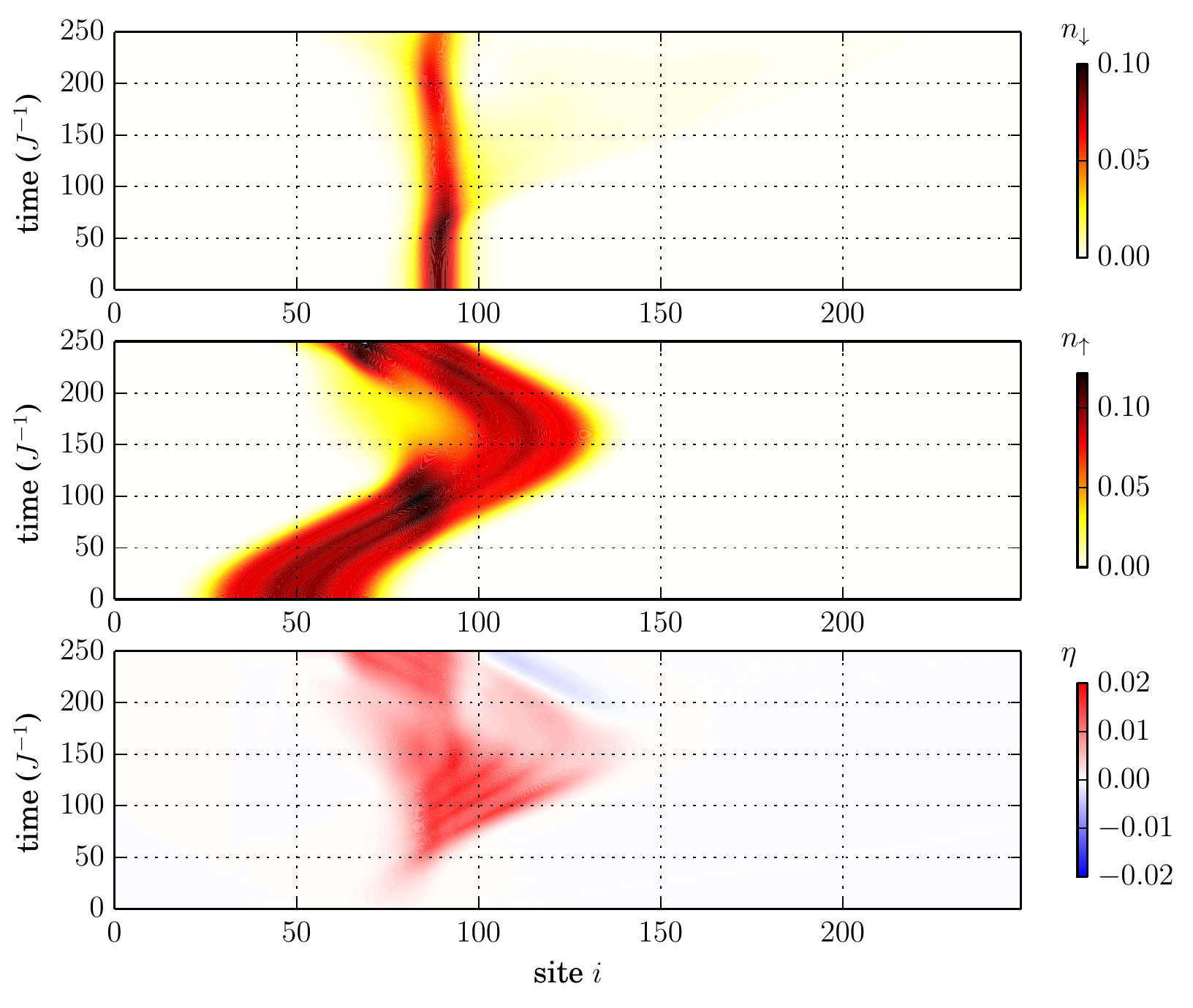}
 \caption{(color online). As in FIG.\ \ref{fig:doublonatt}, but for $U/J = 0.3$.}
 \label{fig:doublonrep}
\end{figure}

The reflected branch of the impurity (see FIG.\ \ref{fig:densities}) propagates as an undisturbed, expanding Gaussian wave packet with a constant propagation velocity of the center of mass.
Intuitively, one would expect this propagation velocity to depend on the ``kick'' given to the majority bosons $(x_\text{B}-x_\text{A})$.
This quantity can be associated with the momentum $k$ in Eq.\ \eqref{eq:scattering}.
Our numerics appear to support this picture for small variations around our choice of $(x_\text{B}-x_\text{A})$, although we cannot perform a systematic analysis of the dependence on $(x_\text{B}-x_\text{A})$ due to numerical constraints.
In addition, we find that for a number of majority bosons $N_\ua = 4$ compared to a single $\ua$-boson, the impurity receives a larger kick, indicated by the steeper slope in FIG.\ \ref{fig:densities}, as it interacts with multiple particles.

\emph{Conclusions.---}
We study the collision of a one-dimensional Bose gas with a weakly trapped impurity.
The probability of the impurity being kicked out of the box potential depends primarily on the \emph{magnitude} of the interaction between the majority component bosons and the impurity, but only weakly on the sign of the interaction.
This is explained in terms of a two-particle scattering picture.

We predict a pair tunneling effect when the interaction energy and the single-particle ground state energy $E_0$ of the impurity are of the same order.
This can be compared to pair tunneling of pairs of attractively interacting distinguishable particles that are prepared in a trap \cite{Zurn2013a,Rontani2012a,Rontani2013a}.
In that case, the interaction leads to a \emph{reduced} tunneling probability of the pair from the trap, as the (negative) interaction energy leads to an effective higher barrier height.
Conversely, here we find an \emph{increased} tunneling probability mediated by the attractive interaction because the energy released when the pair forms is converted to kinetic energy of the pair, allowing it to tunnel outside the box potential.
This feature is broad (see FIG.\ \ref{fig:overlap}b) because the majority component bosons have a large spread in kinetic energy.
Furthermore, our results indicate that the enhanced tunneling peak moves towards higher interactions for deeper box potentials (lower $E_0$), consistent with the pair tunneling picture.

The majority component bosons are found near the Tonks-Girardeau limit $U_\ua/J \rightarrow \infty$.
In this limit, ``fermionization'' occurs, where local observables are identical to a non-interacting spinless gas of fermions.
However, the momentum distribution of the Tonks-Girardeau gas is very different, exhibiting a characteristic $1/q^4$ decay, where $q$ is momentum \cite{Olshanii2003a}.
It would be of interest to see whether this affects the problem considered here.

One aspect that is missing in the picture of Eq.\ \eqref{eq:scattering} is the depth of the box potential $V_\da$.
Indeed, for the impurity to tunnel out of the box potential, it needs an additional energy $\approx E_0$.
This opens up the prospect of using the scheme outlined here to measure the off-shell scattering amplitude (proportional to the two-body $t$-matrix), where the energy is not conserved in the collision.

\emph{Acknowledgments.---}
We thank Th.\ Busch, N.\ T.\ Zinner and M.\ Valiente for useful discussions.
This work was supported by the Academy of Finland through its Centres of Excellence Programme (2012-2017) and under  Project  Nos.  263347,  251748  and  272490, and by the European Research Council (ERC-2013-AdG-340748-CODE).
The numerical results presented in this work have been obtained by using the {\small TDMRG} code developed by S.P.\ in collaboration with Davide Rossini, and by using the {\small LAPACK} \cite{Lapack1999} numerical library.

\bibliographystyle{apsrev4-1}
\bibliography{ref}

\begin{thebibliography}{85}%
\makeatletter
\providecommand \@ifxundefined [1]{%
 \@ifx{#1\undefined}
}%
\providecommand \@ifnum [1]{%
 \ifnum #1\expandafter \@firstoftwo
 \else \expandafter \@secondoftwo
 \fi
}%
\providecommand \@ifx [1]{%
 \ifx #1\expandafter \@firstoftwo
 \else \expandafter \@secondoftwo
 \fi
}%
\providecommand \natexlab [1]{#1}%
\providecommand \enquote  [1]{``#1''}%
\providecommand \bibnamefont  [1]{#1}%
\providecommand \bibfnamefont [1]{#1}%
\providecommand \citenamefont [1]{#1}%
\providecommand \href@noop [0]{\@secondoftwo}%
\providecommand \href [0]{\begingroup \@sanitize@url \@href}%
\providecommand \@href[1]{\@@startlink{#1}\@@href}%
\providecommand \@@href[1]{\endgroup#1\@@endlink}%
\providecommand \@sanitize@url [0]{\catcode `\\12\catcode `\$12\catcode
  `\&12\catcode `\#12\catcode `\^12\catcode `\_12\catcode `\%12\relax}%
\providecommand \@@startlink[1]{}%
\providecommand \@@endlink[0]{}%
\providecommand \url  [0]{\begingroup\@sanitize@url \@url }%
\providecommand \@url [1]{\endgroup\@href {#1}{\urlprefix }}%
\providecommand \urlprefix  [0]{URL }%
\providecommand \Eprint [0]{\href }%
\providecommand \doibase [0]{http://dx.doi.org/}%
\providecommand \selectlanguage [0]{\@gobble}%
\providecommand \bibinfo  [0]{\@secondoftwo}%
\providecommand \bibfield  [0]{\@secondoftwo}%
\providecommand \translation [1]{[#1]}%
\providecommand \BibitemOpen [0]{}%
\providecommand \bibitemStop [0]{}%
\providecommand \bibitemNoStop [0]{.\EOS\space}%
\providecommand \EOS [0]{\spacefactor3000\relax}%
\providecommand \BibitemShut  [1]{\csname bibitem#1\endcsname}%
\let\auto@bib@innerbib\@empty
\bibitem [{\citenamefont {Moritz}\ \emph {et~al.}(2003)\citenamefont {Moritz},
  \citenamefont {St\"oferle}, \citenamefont {K\"ohl},\ and\ \citenamefont
  {Esslinger}}]{Moritz2003a}%
  \BibitemOpen
  \bibfield  {author} {\bibinfo {author} {\bibfnamefont {H.}~\bibnamefont
  {Moritz}}, \bibinfo {author} {\bibfnamefont {T.}~\bibnamefont {St\"oferle}},
  \bibinfo {author} {\bibfnamefont {M.}~\bibnamefont {K\"ohl}}, \ and\ \bibinfo
  {author} {\bibfnamefont {T.}~\bibnamefont {Esslinger}},\ }\href {\doibase
  10.1103/PhysRevLett.91.250402} {\bibfield  {journal} {\bibinfo  {journal}
  {Phys. Rev. Lett.}\ }\textbf {\bibinfo {volume} {91}},\ \bibinfo {pages}
  {250402} (\bibinfo {year} {2003})}\BibitemShut {NoStop}%
\bibitem [{\citenamefont {Kinoshita}\ \emph {et~al.}(2004)\citenamefont
  {Kinoshita}, \citenamefont {Wenger},\ and\ \citenamefont
  {Weiss}}]{Kinoshita2004a}%
  \BibitemOpen
  \bibfield  {author} {\bibinfo {author} {\bibfnamefont {T.}~\bibnamefont
  {Kinoshita}}, \bibinfo {author} {\bibfnamefont {T.}~\bibnamefont {Wenger}}, \
  and\ \bibinfo {author} {\bibfnamefont {D.~S.}\ \bibnamefont {Weiss}},\ }\href
  {\doibase 10.1126/science.1100700} {\bibfield  {journal} {\bibinfo  {journal}
  {Science}\ }\textbf {\bibinfo {volume} {305}},\ \bibinfo {pages} {1125}
  (\bibinfo {year} {2004})}\BibitemShut {NoStop}%
\bibitem [{\citenamefont {Paredes}\ \emph {et~al.}(2004)\citenamefont
  {Paredes}, \citenamefont {Widera}, \citenamefont {Murg}, \citenamefont
  {Mandel}, \citenamefont {F\"olling}, \citenamefont {Cirac}, \citenamefont
  {Shlyapnikov}, \citenamefont {H\"ansch},\ and\ \citenamefont
  {Bloch}}]{Paredes2004a}%
  \BibitemOpen
  \bibfield  {author} {\bibinfo {author} {\bibfnamefont {B.}~\bibnamefont
  {Paredes}}, \bibinfo {author} {\bibfnamefont {A.}~\bibnamefont {Widera}},
  \bibinfo {author} {\bibfnamefont {V.}~\bibnamefont {Murg}}, \bibinfo {author}
  {\bibfnamefont {O.}~\bibnamefont {Mandel}}, \bibinfo {author} {\bibfnamefont
  {S.}~\bibnamefont {F\"olling}}, \bibinfo {author} {\bibfnamefont
  {I.}~\bibnamefont {Cirac}}, \bibinfo {author} {\bibfnamefont {G.~V.}\
  \bibnamefont {Shlyapnikov}}, \bibinfo {author} {\bibfnamefont {T.~W.}\
  \bibnamefont {H\"ansch}}, \ and\ \bibinfo {author} {\bibfnamefont
  {I.}~\bibnamefont {Bloch}},\ }\href {\doibase 10.1038/nature02530} {\bibfield
   {journal} {\bibinfo  {journal} {Nature}\ }\textbf {\bibinfo {volume}
  {429}},\ \bibinfo {pages} {277} (\bibinfo {year} {2004})}\BibitemShut
  {NoStop}%
\bibitem [{\citenamefont {Olshanii}(1998)}]{Olshanii1998a}%
  \BibitemOpen
  \bibfield  {author} {\bibinfo {author} {\bibfnamefont {M.}~\bibnamefont
  {Olshanii}},\ }\href {\doibase 10.1103/PhysRevLett.81.938} {\bibfield
  {journal} {\bibinfo  {journal} {Phys. Rev. Lett.}\ }\textbf {\bibinfo
  {volume} {81}},\ \bibinfo {pages} {938} (\bibinfo {year} {1998})}\BibitemShut
  {NoStop}%
\bibitem [{\citenamefont {Kokkelmans}(2014)}]{Kokkelmans2014a}%
  \BibitemOpen
  \bibfield  {author} {\bibinfo {author} {\bibfnamefont {S.~J. J. M.~F.}\
  \bibnamefont {Kokkelmans}},\ }in\ \href@noop {} {\emph {\bibinfo {booktitle}
  {Quantum gas experiments - exploring many-body states}}},\ \bibinfo {editor}
  {edited by\ \bibinfo {editor} {\bibfnamefont {P.}~\bibnamefont {T\"orm\"a}}\
  and\ \bibinfo {editor} {\bibfnamefont {K.}~\bibnamefont {Sengstock}}}\
  (\bibinfo  {publisher} {Imperial College Press},\ \bibinfo {address}
  {London},\ \bibinfo {year} {2014})\ Chap.~\bibinfo {chapter} {4}\BibitemShut
  {NoStop}%
\bibitem [{\citenamefont {Bergeman}\ \emph {et~al.}(2003)\citenamefont
  {Bergeman}, \citenamefont {Moore},\ and\ \citenamefont
  {Olshanii}}]{Bergeman2003a}%
  \BibitemOpen
  \bibfield  {author} {\bibinfo {author} {\bibfnamefont {T.}~\bibnamefont
  {Bergeman}}, \bibinfo {author} {\bibfnamefont {M.~G.}\ \bibnamefont {Moore}},
  \ and\ \bibinfo {author} {\bibfnamefont {M.}~\bibnamefont {Olshanii}},\
  }\href {\doibase 10.1103/PhysRevLett.91.163201} {\bibfield  {journal}
  {\bibinfo  {journal} {Phys. Rev. Lett.}\ }\textbf {\bibinfo {volume} {91}},\
  \bibinfo {pages} {163201} (\bibinfo {year} {2003})}\BibitemShut {NoStop}%
\bibitem [{\citenamefont {Haller}\ \emph {et~al.}(2010)\citenamefont {Haller},
  \citenamefont {Mark}, \citenamefont {Hart}, \citenamefont {Danzl},
  \citenamefont {Reichs\"ollner}, \citenamefont {Melezhik}, \citenamefont
  {Schmelcher},\ and\ \citenamefont {N\"agerl}}]{Haller2010a}%
  \BibitemOpen
  \bibfield  {author} {\bibinfo {author} {\bibfnamefont {E.}~\bibnamefont
  {Haller}}, \bibinfo {author} {\bibfnamefont {M.~J.}\ \bibnamefont {Mark}},
  \bibinfo {author} {\bibfnamefont {R.}~\bibnamefont {Hart}}, \bibinfo {author}
  {\bibfnamefont {J.~G.}\ \bibnamefont {Danzl}}, \bibinfo {author}
  {\bibfnamefont {L.}~\bibnamefont {Reichs\"ollner}}, \bibinfo {author}
  {\bibfnamefont {V.}~\bibnamefont {Melezhik}}, \bibinfo {author}
  {\bibfnamefont {P.}~\bibnamefont {Schmelcher}}, \ and\ \bibinfo {author}
  {\bibfnamefont {H.-C.}\ \bibnamefont {N\"agerl}},\ }\href {\doibase
  10.1103/PhysRevLett.104.153203} {\bibfield  {journal} {\bibinfo  {journal}
  {Phys. Rev. Lett.}\ }\textbf {\bibinfo {volume} {104}},\ \bibinfo {pages}
  {153203} (\bibinfo {year} {2010})}\BibitemShut {NoStop}%
\bibitem [{\citenamefont {van Amerongen}\ \emph {et~al.}(2008)\citenamefont
  {van Amerongen}, \citenamefont {van Es}, \citenamefont {Wicke}, \citenamefont
  {Kheruntsyan},\ and\ \citenamefont {van Druten}}]{vanAmerongen2008a}%
  \BibitemOpen
  \bibfield  {author} {\bibinfo {author} {\bibfnamefont {A.~H.}\ \bibnamefont
  {van Amerongen}}, \bibinfo {author} {\bibfnamefont {J.~J.~P.}\ \bibnamefont
  {van Es}}, \bibinfo {author} {\bibfnamefont {P.}~\bibnamefont {Wicke}},
  \bibinfo {author} {\bibfnamefont {K.~V.}\ \bibnamefont {Kheruntsyan}}, \ and\
  \bibinfo {author} {\bibfnamefont {N.~J.}\ \bibnamefont {van Druten}},\ }\href
  {\doibase 10.1103/PhysRevLett.100.090402} {\bibfield  {journal} {\bibinfo
  {journal} {Phys. Rev. Lett.}\ }\textbf {\bibinfo {volume} {100}},\ \bibinfo
  {pages} {090402} (\bibinfo {year} {2008})}\BibitemShut {NoStop}%
\bibitem [{\citenamefont {Haller}\ \emph {et~al.}(2009)\citenamefont {Haller},
  \citenamefont {Gustavsson}, \citenamefont {Mark}, \citenamefont {Danzl},
  \citenamefont {Hart}, \citenamefont {Pupillo},\ and\ \citenamefont
  {N\"agerl}}]{Haller2009a}%
  \BibitemOpen
  \bibfield  {author} {\bibinfo {author} {\bibfnamefont {E.}~\bibnamefont
  {Haller}}, \bibinfo {author} {\bibfnamefont {M.}~\bibnamefont {Gustavsson}},
  \bibinfo {author} {\bibfnamefont {M.~J.}\ \bibnamefont {Mark}}, \bibinfo
  {author} {\bibfnamefont {J.~G.}\ \bibnamefont {Danzl}}, \bibinfo {author}
  {\bibfnamefont {R.}~\bibnamefont {Hart}}, \bibinfo {author} {\bibfnamefont
  {G.}~\bibnamefont {Pupillo}}, \ and\ \bibinfo {author} {\bibfnamefont
  {H.-C.}\ \bibnamefont {N\"agerl}},\ }\href {\doibase 10.1126/science.1175850}
  {\bibfield  {journal} {\bibinfo  {journal} {Science}\ }\textbf {\bibinfo
  {volume} {325}},\ \bibinfo {pages} {1224} (\bibinfo {year}
  {2009})}\BibitemShut {NoStop}%
\bibitem [{\citenamefont {Fabbri}\ \emph {et~al.}(2015)\citenamefont {Fabbri},
  \citenamefont {Panfil}, \citenamefont {Cl\'ement}, \citenamefont {Fallani},
  \citenamefont {Inguscio}, \citenamefont {Fort},\ and\ \citenamefont
  {Caux}}]{Fabbri2015a}%
  \BibitemOpen
  \bibfield  {author} {\bibinfo {author} {\bibfnamefont {N.}~\bibnamefont
  {Fabbri}}, \bibinfo {author} {\bibfnamefont {M.}~\bibnamefont {Panfil}},
  \bibinfo {author} {\bibfnamefont {D.}~\bibnamefont {Cl\'ement}}, \bibinfo
  {author} {\bibfnamefont {L.}~\bibnamefont {Fallani}}, \bibinfo {author}
  {\bibfnamefont {M.}~\bibnamefont {Inguscio}}, \bibinfo {author}
  {\bibfnamefont {C.}~\bibnamefont {Fort}}, \ and\ \bibinfo {author}
  {\bibfnamefont {J.-S.}\ \bibnamefont {Caux}},\ }\href {\doibase
  10.1103/PhysRevA.91.043617} {\bibfield  {journal} {\bibinfo  {journal} {Phys.
  Rev. A}\ }\textbf {\bibinfo {volume} {91}},\ \bibinfo {pages} {043617}
  (\bibinfo {year} {2015})}\BibitemShut {NoStop}%
\bibitem [{\citenamefont {Kinoshita}\ \emph {et~al.}(2006)\citenamefont
  {Kinoshita}, \citenamefont {Wenger},\ and\ \citenamefont
  {Weiss}}]{Kinoshita2006a}%
  \BibitemOpen
  \bibfield  {author} {\bibinfo {author} {\bibfnamefont {T.}~\bibnamefont
  {Kinoshita}}, \bibinfo {author} {\bibfnamefont {T.}~\bibnamefont {Wenger}}, \
  and\ \bibinfo {author} {\bibfnamefont {D.~S.}\ \bibnamefont {Weiss}},\ }\href
  {\doibase 10.1038/nature04693} {\bibfield  {journal} {\bibinfo  {journal}
  {Nature}\ }\textbf {\bibinfo {volume} {440}},\ \bibinfo {pages} {900}
  (\bibinfo {year} {2006})}\BibitemShut {NoStop}%
\bibitem [{\citenamefont {Hofferberth}\ \emph {et~al.}(2007)\citenamefont
  {Hofferberth}, \citenamefont {Lesanovsky}, \citenamefont {Fischer},
  \citenamefont {Schumm},\ and\ \citenamefont
  {Schmiedmayer}}]{Hofferberth2007a}%
  \BibitemOpen
  \bibfield  {author} {\bibinfo {author} {\bibfnamefont {S.}~\bibnamefont
  {Hofferberth}}, \bibinfo {author} {\bibfnamefont {I.}~\bibnamefont
  {Lesanovsky}}, \bibinfo {author} {\bibfnamefont {B.}~\bibnamefont {Fischer}},
  \bibinfo {author} {\bibfnamefont {T.}~\bibnamefont {Schumm}}, \ and\ \bibinfo
  {author} {\bibfnamefont {J.}~\bibnamefont {Schmiedmayer}},\ }\href@noop {}
  {\bibfield  {journal} {\bibinfo  {journal} {Nature}\ }\textbf {\bibinfo
  {volume} {449}},\ \bibinfo {pages} {324} (\bibinfo {year}
  {2007})}\BibitemShut {NoStop}%
\bibitem [{\citenamefont {Gring}\ \emph {et~al.}(2012)\citenamefont {Gring},
  \citenamefont {Kuhnert}, \citenamefont {Langen}, \citenamefont {Kitagawa},
  \citenamefont {Rauer}, \citenamefont {Schreitl}, \citenamefont {Mazets},
  \citenamefont {Smith}, \citenamefont {Demler},\ and\ \citenamefont
  {Schmiedmayer}}]{Gring2012a}%
  \BibitemOpen
  \bibfield  {author} {\bibinfo {author} {\bibfnamefont {M.}~\bibnamefont
  {Gring}}, \bibinfo {author} {\bibfnamefont {M.}~\bibnamefont {Kuhnert}},
  \bibinfo {author} {\bibfnamefont {T.}~\bibnamefont {Langen}}, \bibinfo
  {author} {\bibfnamefont {T.}~\bibnamefont {Kitagawa}}, \bibinfo {author}
  {\bibfnamefont {B.}~\bibnamefont {Rauer}}, \bibinfo {author} {\bibfnamefont
  {M.}~\bibnamefont {Schreitl}}, \bibinfo {author} {\bibfnamefont
  {I.}~\bibnamefont {Mazets}}, \bibinfo {author} {\bibfnamefont {D.~A.}\
  \bibnamefont {Smith}}, \bibinfo {author} {\bibfnamefont {E.}~\bibnamefont
  {Demler}}, \ and\ \bibinfo {author} {\bibfnamefont {J.}~\bibnamefont
  {Schmiedmayer}},\ }\href {\doibase 10.1126/science.1224953} {\bibfield
  {journal} {\bibinfo  {journal} {Science}\ }\textbf {\bibinfo {volume}
  {337}},\ \bibinfo {pages} {1318} (\bibinfo {year} {2012})}\BibitemShut
  {NoStop}%
\bibitem [{\citenamefont {Trotzky}\ \emph {et~al.}(2012)\citenamefont
  {Trotzky}, \citenamefont {Chen}, \citenamefont {Flesch}, \citenamefont
  {McCulloch}, \citenamefont {Schollw\"ock}, \citenamefont {Eisert},\ and\
  \citenamefont {Bloch}}]{Trotzky2012a}%
  \BibitemOpen
  \bibfield  {author} {\bibinfo {author} {\bibfnamefont {S.}~\bibnamefont
  {Trotzky}}, \bibinfo {author} {\bibfnamefont {Y.-A.}\ \bibnamefont {Chen}},
  \bibinfo {author} {\bibfnamefont {A.}~\bibnamefont {Flesch}}, \bibinfo
  {author} {\bibfnamefont {I.~P.}\ \bibnamefont {McCulloch}}, \bibinfo {author}
  {\bibfnamefont {U.}~\bibnamefont {Schollw\"ock}}, \bibinfo {author}
  {\bibfnamefont {J.}~\bibnamefont {Eisert}}, \ and\ \bibinfo {author}
  {\bibfnamefont {I.}~\bibnamefont {Bloch}},\ }\href@noop {} {\bibfield
  {journal} {\bibinfo  {journal} {Nature Phys.}\ }\textbf {\bibinfo {volume}
  {8}},\ \bibinfo {pages} {325} (\bibinfo {year} {2012})}\BibitemShut {NoStop}%
\bibitem [{\citenamefont {Ronzheimer}\ \emph {et~al.}(2013)\citenamefont
  {Ronzheimer}, \citenamefont {Schreiber}, \citenamefont {Braun}, \citenamefont
  {Hodgman}, \citenamefont {Langer}, \citenamefont {McCulloch}, \citenamefont
  {Heidrich-Meisner}, \citenamefont {Bloch},\ and\ \citenamefont
  {Schneider}}]{Ronzheimer2013a}%
  \BibitemOpen
  \bibfield  {author} {\bibinfo {author} {\bibfnamefont {J.~P.}\ \bibnamefont
  {Ronzheimer}}, \bibinfo {author} {\bibfnamefont {M.}~\bibnamefont
  {Schreiber}}, \bibinfo {author} {\bibfnamefont {S.}~\bibnamefont {Braun}},
  \bibinfo {author} {\bibfnamefont {S.~S.}\ \bibnamefont {Hodgman}}, \bibinfo
  {author} {\bibfnamefont {S.}~\bibnamefont {Langer}}, \bibinfo {author}
  {\bibfnamefont {I.~P.}\ \bibnamefont {McCulloch}}, \bibinfo {author}
  {\bibfnamefont {F.}~\bibnamefont {Heidrich-Meisner}}, \bibinfo {author}
  {\bibfnamefont {I.}~\bibnamefont {Bloch}}, \ and\ \bibinfo {author}
  {\bibfnamefont {U.}~\bibnamefont {Schneider}},\ }\href {\doibase
  10.1103/PhysRevLett.110.205301} {\bibfield  {journal} {\bibinfo  {journal}
  {Phys. Rev. Lett.}\ }\textbf {\bibinfo {volume} {110}},\ \bibinfo {pages}
  {205301} (\bibinfo {year} {2013})}\BibitemShut {NoStop}%
\bibitem [{\citenamefont {Langen}\ \emph {et~al.}(2013)\citenamefont {Langen},
  \citenamefont {Geiger}, \citenamefont {Kuhnert}, \citenamefont {Rauer},\ and\
  \citenamefont {Scheidmaier}}]{Langen2013a}%
  \BibitemOpen
  \bibfield  {author} {\bibinfo {author} {\bibfnamefont {T.}~\bibnamefont
  {Langen}}, \bibinfo {author} {\bibfnamefont {R.}~\bibnamefont {Geiger}},
  \bibinfo {author} {\bibfnamefont {M.}~\bibnamefont {Kuhnert}}, \bibinfo
  {author} {\bibfnamefont {B.}~\bibnamefont {Rauer}}, \ and\ \bibinfo {author}
  {\bibfnamefont {J.}~\bibnamefont {Scheidmaier}},\ }\href {\doibase
  10.1038/NPHYS2739} {\bibfield  {journal} {\bibinfo  {journal} {Nature Phys.}\
  }\textbf {\bibinfo {volume} {9}},\ \bibinfo {pages} {640} (\bibinfo {year}
  {2013})}\BibitemShut {NoStop}%
\bibitem [{\citenamefont {Palzer}\ \emph {et~al.}(2009)\citenamefont {Palzer},
  \citenamefont {Zipkes}, \citenamefont {Sias},\ and\ \citenamefont
  {K\"ohl}}]{Palzer2009a}%
  \BibitemOpen
  \bibfield  {author} {\bibinfo {author} {\bibfnamefont {S.}~\bibnamefont
  {Palzer}}, \bibinfo {author} {\bibfnamefont {C.}~\bibnamefont {Zipkes}},
  \bibinfo {author} {\bibfnamefont {C.}~\bibnamefont {Sias}}, \ and\ \bibinfo
  {author} {\bibfnamefont {M.}~\bibnamefont {K\"ohl}},\ }\href {\doibase
  10.1103/PhysRevLett.103.150601} {\bibfield  {journal} {\bibinfo  {journal}
  {Phys. Rev. Lett.}\ }\textbf {\bibinfo {volume} {103}},\ \bibinfo {pages}
  {150601} (\bibinfo {year} {2009})}\BibitemShut {NoStop}%
\bibitem [{\citenamefont {Spethmann}\ \emph {et~al.}(2012)\citenamefont
  {Spethmann}, \citenamefont {Kindermann}, \citenamefont {John}, \citenamefont
  {Weber}, \citenamefont {Meschede},\ and\ \citenamefont
  {Widera}}]{Spethmann2012a}%
  \BibitemOpen
  \bibfield  {author} {\bibinfo {author} {\bibfnamefont {N.}~\bibnamefont
  {Spethmann}}, \bibinfo {author} {\bibfnamefont {F.}~\bibnamefont
  {Kindermann}}, \bibinfo {author} {\bibfnamefont {S.}~\bibnamefont {John}},
  \bibinfo {author} {\bibfnamefont {C.}~\bibnamefont {Weber}}, \bibinfo
  {author} {\bibfnamefont {D.}~\bibnamefont {Meschede}}, \ and\ \bibinfo
  {author} {\bibfnamefont {A.}~\bibnamefont {Widera}},\ }\href {\doibase
  10.1103/PhysRevLett.109.235301} {\bibfield  {journal} {\bibinfo  {journal}
  {Phys. Rev. Lett.}\ }\textbf {\bibinfo {volume} {109}},\ \bibinfo {pages}
  {235301} (\bibinfo {year} {2012})}\BibitemShut {NoStop}%
\bibitem [{\citenamefont {Catani}\ \emph {et~al.}(2012)\citenamefont {Catani},
  \citenamefont {Lamporesi}, \citenamefont {Naik}, \citenamefont {Gring},
  \citenamefont {Inguscio}, \citenamefont {Minardi}, \citenamefont {Kantian},\
  and\ \citenamefont {Giamarchi}}]{Catani2012a}%
  \BibitemOpen
  \bibfield  {author} {\bibinfo {author} {\bibfnamefont {J.}~\bibnamefont
  {Catani}}, \bibinfo {author} {\bibfnamefont {G.}~\bibnamefont {Lamporesi}},
  \bibinfo {author} {\bibfnamefont {D.}~\bibnamefont {Naik}}, \bibinfo {author}
  {\bibfnamefont {M.}~\bibnamefont {Gring}}, \bibinfo {author} {\bibfnamefont
  {M.}~\bibnamefont {Inguscio}}, \bibinfo {author} {\bibfnamefont
  {F.}~\bibnamefont {Minardi}}, \bibinfo {author} {\bibfnamefont
  {A.}~\bibnamefont {Kantian}}, \ and\ \bibinfo {author} {\bibfnamefont
  {T.}~\bibnamefont {Giamarchi}},\ }\href {\doibase 10.1103/PhysRevA.85.023623}
  {\bibfield  {journal} {\bibinfo  {journal} {Phys. Rev. A}\ }\textbf {\bibinfo
  {volume} {85}},\ \bibinfo {pages} {023623} (\bibinfo {year}
  {2012})}\BibitemShut {NoStop}%
\bibitem [{\citenamefont {Fukuhara}\ \emph {et~al.}(2013)\citenamefont
  {Fukuhara}, \citenamefont {Kantian}, \citenamefont {Endres}, \citenamefont
  {Cheneau}, \citenamefont {Schau{\ss}}, \citenamefont {Hild}, \citenamefont
  {Bellem}, \citenamefont {Schollw\"ock}, \citenamefont {Giamarchi},
  \citenamefont {Gross}, \citenamefont {Bloch},\ and\ \citenamefont
  {Kuhr}}]{Fukuhara2013a}%
  \BibitemOpen
  \bibfield  {author} {\bibinfo {author} {\bibfnamefont {T.}~\bibnamefont
  {Fukuhara}}, \bibinfo {author} {\bibfnamefont {A.}~\bibnamefont {Kantian}},
  \bibinfo {author} {\bibfnamefont {M.}~\bibnamefont {Endres}}, \bibinfo
  {author} {\bibfnamefont {M.}~\bibnamefont {Cheneau}}, \bibinfo {author}
  {\bibfnamefont {P.}~\bibnamefont {Schau{\ss}}}, \bibinfo {author}
  {\bibfnamefont {S.}~\bibnamefont {Hild}}, \bibinfo {author} {\bibfnamefont
  {D.}~\bibnamefont {Bellem}}, \bibinfo {author} {\bibfnamefont
  {U.}~\bibnamefont {Schollw\"ock}}, \bibinfo {author} {\bibfnamefont
  {T.}~\bibnamefont {Giamarchi}}, \bibinfo {author} {\bibfnamefont
  {C.}~\bibnamefont {Gross}}, \bibinfo {author} {\bibfnamefont
  {I.}~\bibnamefont {Bloch}}, \ and\ \bibinfo {author} {\bibfnamefont
  {S.}~\bibnamefont {Kuhr}},\ }\href {\doibase 10.1038/nphys2561} {\bibfield
  {journal} {\bibinfo  {journal} {Nature Phys.}\ }\textbf {\bibinfo {volume}
  {9}},\ \bibinfo {pages} {235} (\bibinfo {year} {2013})}\BibitemShut {NoStop}%
\bibitem [{\citenamefont {Z\"urn}\ \emph {et~al.}(2012)\citenamefont {Z\"urn},
  \citenamefont {Serwane}, \citenamefont {Lompe}, \citenamefont {Wenz},
  \citenamefont {Ries}, \citenamefont {Bohn},\ and\ \citenamefont
  {Jochim}}]{Zurn2012a}%
  \BibitemOpen
  \bibfield  {author} {\bibinfo {author} {\bibfnamefont {G.}~\bibnamefont
  {Z\"urn}}, \bibinfo {author} {\bibfnamefont {F.}~\bibnamefont {Serwane}},
  \bibinfo {author} {\bibfnamefont {T.}~\bibnamefont {Lompe}}, \bibinfo
  {author} {\bibfnamefont {A.~N.}\ \bibnamefont {Wenz}}, \bibinfo {author}
  {\bibfnamefont {M.~G.}\ \bibnamefont {Ries}}, \bibinfo {author}
  {\bibfnamefont {J.~E.}\ \bibnamefont {Bohn}}, \ and\ \bibinfo {author}
  {\bibfnamefont {S.}~\bibnamefont {Jochim}},\ }\href {\doibase
  10.1103/PhysRevLett.108.075303} {\bibfield  {journal} {\bibinfo  {journal}
  {Phys. Rev. Lett.}\ }\textbf {\bibinfo {volume} {108}},\ \bibinfo {pages}
  {075303} (\bibinfo {year} {2012})}\BibitemShut {NoStop}%
\bibitem [{\citenamefont {Wenz}\ \emph {et~al.}(2013)\citenamefont {Wenz},
  \citenamefont {Z\"urn}, \citenamefont {Murmann}, \citenamefont {Brouzos},
  \citenamefont {Lompe},\ and\ \citenamefont {Jochim}}]{Wenz2013a}%
  \BibitemOpen
  \bibfield  {author} {\bibinfo {author} {\bibfnamefont {A.~N.}\ \bibnamefont
  {Wenz}}, \bibinfo {author} {\bibfnamefont {G.}~\bibnamefont {Z\"urn}},
  \bibinfo {author} {\bibfnamefont {S.}~\bibnamefont {Murmann}}, \bibinfo
  {author} {\bibfnamefont {I.}~\bibnamefont {Brouzos}}, \bibinfo {author}
  {\bibfnamefont {T.}~\bibnamefont {Lompe}}, \ and\ \bibinfo {author}
  {\bibfnamefont {S.}~\bibnamefont {Jochim}},\ }\href {\doibase
  10.1126/science.1240516} {\bibfield  {journal} {\bibinfo  {journal}
  {Science}\ }\textbf {\bibinfo {volume} {342}},\ \bibinfo {pages} {457}
  (\bibinfo {year} {2013})}\BibitemShut {NoStop}%
\bibitem [{\citenamefont {Z\"urn}\ \emph {et~al.}(2013)\citenamefont {Z\"urn},
  \citenamefont {Wenz}, \citenamefont {Murmann}, \citenamefont {Bergschneider},
  \citenamefont {Lompe},\ and\ \citenamefont {Jochim}}]{Zurn2013a}%
  \BibitemOpen
  \bibfield  {author} {\bibinfo {author} {\bibfnamefont {G.}~\bibnamefont
  {Z\"urn}}, \bibinfo {author} {\bibfnamefont {A.~N.}\ \bibnamefont {Wenz}},
  \bibinfo {author} {\bibfnamefont {S.}~\bibnamefont {Murmann}}, \bibinfo
  {author} {\bibfnamefont {A.}~\bibnamefont {Bergschneider}}, \bibinfo {author}
  {\bibfnamefont {T.}~\bibnamefont {Lompe}}, \ and\ \bibinfo {author}
  {\bibfnamefont {S.}~\bibnamefont {Jochim}},\ }\href {\doibase
  10.1103/PhysRevLett.111.175302} {\bibfield  {journal} {\bibinfo  {journal}
  {Phys. Rev. Lett.}\ }\textbf {\bibinfo {volume} {111}},\ \bibinfo {pages}
  {175302} (\bibinfo {year} {2013})}\BibitemShut {NoStop}%
\bibitem [{\citenamefont {Giamarchi}(2003)}]{Giamarchi2003a}%
  \BibitemOpen
  \bibfield  {author} {\bibinfo {author} {\bibfnamefont {T.}~\bibnamefont
  {Giamarchi}},\ }\href@noop {} {\emph {\bibinfo {title} {Quantum Physics in
  One Dimension}}}\ (\bibinfo  {publisher} {Clarendon Press},\ \bibinfo
  {address} {Oxford},\ \bibinfo {year} {2003})\BibitemShut {NoStop}%
\bibitem [{\citenamefont {Cazalilla}\ \emph {et~al.}(2011)\citenamefont
  {Cazalilla}, \citenamefont {Citro}, \citenamefont {Giamarchi}, \citenamefont
  {Orignac},\ and\ \citenamefont {Rigol}}]{Cazalilla2011a}%
  \BibitemOpen
  \bibfield  {author} {\bibinfo {author} {\bibfnamefont {M.~A.}\ \bibnamefont
  {Cazalilla}}, \bibinfo {author} {\bibfnamefont {R.}~\bibnamefont {Citro}},
  \bibinfo {author} {\bibfnamefont {T.}~\bibnamefont {Giamarchi}}, \bibinfo
  {author} {\bibfnamefont {E.}~\bibnamefont {Orignac}}, \ and\ \bibinfo
  {author} {\bibfnamefont {M.}~\bibnamefont {Rigol}},\ }\href {\doibase
  10.1103/RevModPhys.83.1405} {\bibfield  {journal} {\bibinfo  {journal} {Rev.
  Mod. Phys.}\ }\textbf {\bibinfo {volume} {83}},\ \bibinfo {pages} {1405}
  (\bibinfo {year} {2011})}\BibitemShut {NoStop}%
\bibitem [{\citenamefont {Imambekov}\ \emph {et~al.}(2012)\citenamefont
  {Imambekov}, \citenamefont {Schmidt},\ and\ \citenamefont
  {Glazman}}]{Imambekov2012a}%
  \BibitemOpen
  \bibfield  {author} {\bibinfo {author} {\bibfnamefont {A.}~\bibnamefont
  {Imambekov}}, \bibinfo {author} {\bibfnamefont {T.~L.}\ \bibnamefont
  {Schmidt}}, \ and\ \bibinfo {author} {\bibfnamefont {L.~I.}\ \bibnamefont
  {Glazman}},\ }\href {\doibase 10.1103/RevModPhys.84.1253} {\bibfield
  {journal} {\bibinfo  {journal} {Rev. Mod. Phys.}\ }\textbf {\bibinfo {volume}
  {84}},\ \bibinfo {pages} {1253} (\bibinfo {year} {2012})}\BibitemShut
  {NoStop}%
\bibitem [{\citenamefont {Guan}\ \emph {et~al.}(2013)\citenamefont {Guan},
  \citenamefont {Batchelor},\ and\ \citenamefont {Lee}}]{Guan2013a}%
  \BibitemOpen
  \bibfield  {author} {\bibinfo {author} {\bibfnamefont {X.-W.}\ \bibnamefont
  {Guan}}, \bibinfo {author} {\bibfnamefont {M.~T.}\ \bibnamefont {Batchelor}},
  \ and\ \bibinfo {author} {\bibfnamefont {C.}~\bibnamefont {Lee}},\ }\href
  {\doibase 10.1103/RevModPhys.85.1633} {\bibfield  {journal} {\bibinfo
  {journal} {Rev. Mod. Phys.}\ }\textbf {\bibinfo {volume} {85}},\ \bibinfo
  {pages} {1633} (\bibinfo {year} {2013})}\BibitemShut {NoStop}%
\bibitem [{\citenamefont {Schollw\"ock}(2011)}]{Schollwock2011a}%
  \BibitemOpen
  \bibfield  {author} {\bibinfo {author} {\bibfnamefont {U.}~\bibnamefont
  {Schollw\"ock}},\ }\href {\doibase
  http://dx.doi.org/10.1016/j.aop.2010.09.012} {\bibfield  {journal} {\bibinfo
  {journal} {Ann. Phys.}\ }\textbf {\bibinfo {volume} {326}},\ \bibinfo {pages}
  {96 } (\bibinfo {year} {2011})}\BibitemShut {NoStop}%
\bibitem [{\citenamefont {Vidal}(2003)}]{Vidal2003a}%
  \BibitemOpen
  \bibfield  {author} {\bibinfo {author} {\bibfnamefont {G.}~\bibnamefont
  {Vidal}},\ }\href {\doibase 10.1103/PhysRevLett.91.147902} {\bibfield
  {journal} {\bibinfo  {journal} {Phys. Rev. Lett.}\ }\textbf {\bibinfo
  {volume} {91}},\ \bibinfo {pages} {147902} (\bibinfo {year}
  {2003})}\BibitemShut {NoStop}%
\bibitem [{\citenamefont {Zhang}\ and\ \citenamefont
  {Dong}(2010)}]{Zhang2010a}%
  \BibitemOpen
  \bibfield  {author} {\bibinfo {author} {\bibfnamefont {J.~M.}\ \bibnamefont
  {Zhang}}\ and\ \bibinfo {author} {\bibfnamefont {R.~X.}\ \bibnamefont
  {Dong}},\ }\href {http://stacks.iop.org/0143-0807/31/i=3/a=016} {\bibfield
  {journal} {\bibinfo  {journal} {Eur. J. Phys.}\ }\textbf {\bibinfo {volume}
  {31}},\ \bibinfo {pages} {591} (\bibinfo {year} {2010})}\BibitemShut
  {NoStop}%
\bibitem [{\citenamefont {Johnson}\ \emph {et~al.}(2012)\citenamefont
  {Johnson}, \citenamefont {Bruderer}, \citenamefont {Cai}, \citenamefont
  {Clark}, \citenamefont {Bao},\ and\ \citenamefont {Jaksch}}]{Johnson2012a}%
  \BibitemOpen
  \bibfield  {author} {\bibinfo {author} {\bibfnamefont {T.~H.}\ \bibnamefont
  {Johnson}}, \bibinfo {author} {\bibfnamefont {M.}~\bibnamefont {Bruderer}},
  \bibinfo {author} {\bibfnamefont {Y.}~\bibnamefont {Cai}}, \bibinfo {author}
  {\bibfnamefont {S.~R.}\ \bibnamefont {Clark}}, \bibinfo {author}
  {\bibfnamefont {W.}~\bibnamefont {Bao}}, \ and\ \bibinfo {author}
  {\bibfnamefont {D.}~\bibnamefont {Jaksch}},\ }\href
  {http://stacks.iop.org/0295-5075/98/i=2/a=26001} {\bibfield  {journal}
  {\bibinfo  {journal} {Europhys. Lett.}\ }\textbf {\bibinfo {volume} {98}},\
  \bibinfo {pages} {26001} (\bibinfo {year} {2012})}\BibitemShut {NoStop}%
\bibitem [{\citenamefont {Knap}\ \emph {et~al.}(2012)\citenamefont {Knap},
  \citenamefont {Shashi}, \citenamefont {Nishida}, \citenamefont {Imambekov},
  \citenamefont {Abanin},\ and\ \citenamefont {Demler}}]{Knap2012a}%
  \BibitemOpen
  \bibfield  {author} {\bibinfo {author} {\bibfnamefont {M.}~\bibnamefont
  {Knap}}, \bibinfo {author} {\bibfnamefont {A.}~\bibnamefont {Shashi}},
  \bibinfo {author} {\bibfnamefont {Y.}~\bibnamefont {Nishida}}, \bibinfo
  {author} {\bibfnamefont {A.}~\bibnamefont {Imambekov}}, \bibinfo {author}
  {\bibfnamefont {D.~A.}\ \bibnamefont {Abanin}}, \ and\ \bibinfo {author}
  {\bibfnamefont {E.}~\bibnamefont {Demler}},\ }\href {\doibase
  10.1103/PhysRevX.2.041020} {\bibfield  {journal} {\bibinfo  {journal} {Phys.
  Rev. X}\ }\textbf {\bibinfo {volume} {2}},\ \bibinfo {pages} {041020}
  (\bibinfo {year} {2012})}\BibitemShut {NoStop}%
\bibitem [{\citenamefont {Schecter}\ \emph {et~al.}(2012)\citenamefont
  {Schecter}, \citenamefont {Kamenev}, \citenamefont {Gangardt},\ and\
  \citenamefont {Lamacraft}}]{Schecter2012a}%
  \BibitemOpen
  \bibfield  {author} {\bibinfo {author} {\bibfnamefont {M.}~\bibnamefont
  {Schecter}}, \bibinfo {author} {\bibfnamefont {A.}~\bibnamefont {Kamenev}},
  \bibinfo {author} {\bibfnamefont {D.~M.}\ \bibnamefont {Gangardt}}, \ and\
  \bibinfo {author} {\bibfnamefont {A.}~\bibnamefont {Lamacraft}},\ }\href
  {\doibase 10.1103/PhysRevLett.108.207001} {\bibfield  {journal} {\bibinfo
  {journal} {Phys. Rev. Lett.}\ }\textbf {\bibinfo {volume} {108}},\ \bibinfo
  {pages} {207001} (\bibinfo {year} {2012})}\BibitemShut {NoStop}%
\bibitem [{\citenamefont {Mathy}\ \emph {et~al.}(2012)\citenamefont {Mathy},
  \citenamefont {Zvonarev},\ and\ \citenamefont {Demler}}]{Mathy2012a}%
  \BibitemOpen
  \bibfield  {author} {\bibinfo {author} {\bibfnamefont {C.~J.~M.}\
  \bibnamefont {Mathy}}, \bibinfo {author} {\bibfnamefont {M.~B.}\ \bibnamefont
  {Zvonarev}}, \ and\ \bibinfo {author} {\bibfnamefont {E.}~\bibnamefont
  {Demler}},\ }\href {\doibase 10.1038/nphys2455} {\bibfield  {journal}
  {\bibinfo  {journal} {Nature Phys.}\ }\textbf {\bibinfo {volume} {8}},\
  \bibinfo {pages} {881} (\bibinfo {year} {2012})}\BibitemShut {NoStop}%
\bibitem [{\citenamefont {Bonart}\ and\ \citenamefont
  {Cugliandolo}(2012)}]{Bonart2012a}%
  \BibitemOpen
  \bibfield  {author} {\bibinfo {author} {\bibfnamefont {J.}~\bibnamefont
  {Bonart}}\ and\ \bibinfo {author} {\bibfnamefont {L.~F.}\ \bibnamefont
  {Cugliandolo}},\ }\href {\doibase 10.1103/PhysRevA.86.023636} {\bibfield
  {journal} {\bibinfo  {journal} {Phys. Rev. A}\ }\textbf {\bibinfo {volume}
  {86}},\ \bibinfo {pages} {023636} (\bibinfo {year} {2012})}\BibitemShut
  {NoStop}%
\bibitem [{\citenamefont {Massel}\ \emph {et~al.}(2013)\citenamefont {Massel},
  \citenamefont {Kantian}, \citenamefont {Daley}, \citenamefont {Giamarchi},\
  and\ \citenamefont {T\"orm\"a}}]{Massel2013a}%
  \BibitemOpen
  \bibfield  {author} {\bibinfo {author} {\bibfnamefont {F.}~\bibnamefont
  {Massel}}, \bibinfo {author} {\bibfnamefont {A.}~\bibnamefont {Kantian}},
  \bibinfo {author} {\bibfnamefont {A.~J.}\ \bibnamefont {Daley}}, \bibinfo
  {author} {\bibfnamefont {T.}~\bibnamefont {Giamarchi}}, \ and\ \bibinfo
  {author} {\bibfnamefont {P.}~\bibnamefont {T\"orm\"a}},\ }\href@noop {}
  {\bibfield  {journal} {\bibinfo  {journal} {New J. Phys.}\ }\textbf {\bibinfo
  {volume} {15}},\ \bibinfo {pages} {045018} (\bibinfo {year}
  {2013})}\BibitemShut {NoStop}%
\bibitem [{\citenamefont {Sindona}\ \emph {et~al.}(2013)\citenamefont
  {Sindona}, \citenamefont {Goold}, \citenamefont {Lo~Gullo}, \citenamefont
  {Lorenzo},\ and\ \citenamefont {Plastina}}]{Sindona2013a}%
  \BibitemOpen
  \bibfield  {author} {\bibinfo {author} {\bibfnamefont {A.}~\bibnamefont
  {Sindona}}, \bibinfo {author} {\bibfnamefont {J.}~\bibnamefont {Goold}},
  \bibinfo {author} {\bibfnamefont {N.}~\bibnamefont {Lo~Gullo}}, \bibinfo
  {author} {\bibfnamefont {S.}~\bibnamefont {Lorenzo}}, \ and\ \bibinfo
  {author} {\bibfnamefont {F.}~\bibnamefont {Plastina}},\ }\href {\doibase
  10.1103/PhysRevLett.111.165303} {\bibfield  {journal} {\bibinfo  {journal}
  {Phys. Rev. Lett.}\ }\textbf {\bibinfo {volume} {111}},\ \bibinfo {pages}
  {165303} (\bibinfo {year} {2013})}\BibitemShut {NoStop}%
\bibitem [{\citenamefont {Peotta}\ \emph {et~al.}(2013)\citenamefont {Peotta},
  \citenamefont {Rossini}, \citenamefont {Polini}, \citenamefont {Minardi},\
  and\ \citenamefont {Fazio}}]{Peotta2013a}%
  \BibitemOpen
  \bibfield  {author} {\bibinfo {author} {\bibfnamefont {S.}~\bibnamefont
  {Peotta}}, \bibinfo {author} {\bibfnamefont {D.}~\bibnamefont {Rossini}},
  \bibinfo {author} {\bibfnamefont {M.}~\bibnamefont {Polini}}, \bibinfo
  {author} {\bibfnamefont {F.}~\bibnamefont {Minardi}}, \ and\ \bibinfo
  {author} {\bibfnamefont {R.}~\bibnamefont {Fazio}},\ }\href {\doibase
  10.1103/PhysRevLett.110.015302} {\bibfield  {journal} {\bibinfo  {journal}
  {Phys. Rev. Lett.}\ }\textbf {\bibinfo {volume} {110}},\ \bibinfo {pages}
  {015302} (\bibinfo {year} {2013})}\BibitemShut {NoStop}%
\bibitem [{\citenamefont {Kantian}\ \emph {et~al.}(2014)\citenamefont
  {Kantian}, \citenamefont {Schollw\"ock},\ and\ \citenamefont
  {Giamarchi}}]{Kantian2014a}%
  \BibitemOpen
  \bibfield  {author} {\bibinfo {author} {\bibfnamefont {A.}~\bibnamefont
  {Kantian}}, \bibinfo {author} {\bibfnamefont {U.}~\bibnamefont
  {Schollw\"ock}}, \ and\ \bibinfo {author} {\bibfnamefont {T.}~\bibnamefont
  {Giamarchi}},\ }\href {\doibase 10.1103/PhysRevLett.113.070601} {\bibfield
  {journal} {\bibinfo  {journal} {Phys. Rev. Lett.}\ }\textbf {\bibinfo
  {volume} {113}},\ \bibinfo {pages} {070601} (\bibinfo {year}
  {2014})}\BibitemShut {NoStop}%
\bibitem [{\citenamefont {Burovski}\ \emph {et~al.}(2014)\citenamefont
  {Burovski}, \citenamefont {Cheianov}, \citenamefont {Gamayun},\ and\
  \citenamefont {Lychkovskiy}}]{Burovski2014a}%
  \BibitemOpen
  \bibfield  {author} {\bibinfo {author} {\bibfnamefont {E.}~\bibnamefont
  {Burovski}}, \bibinfo {author} {\bibfnamefont {V.}~\bibnamefont {Cheianov}},
  \bibinfo {author} {\bibfnamefont {O.}~\bibnamefont {Gamayun}}, \ and\
  \bibinfo {author} {\bibfnamefont {O.}~\bibnamefont {Lychkovskiy}},\ }\href
  {\doibase 10.1103/PhysRevA.89.041601} {\bibfield  {journal} {\bibinfo
  {journal} {Phys. Rev. A}\ }\textbf {\bibinfo {volume} {89}},\ \bibinfo
  {pages} {041601} (\bibinfo {year} {2014})}\BibitemShut {NoStop}%
\bibitem [{\citenamefont {Visuri}\ \emph {et~al.}(2014)\citenamefont {Visuri},
  \citenamefont {Kim}, \citenamefont {Kinnunen}, \citenamefont {Massel},\ and\
  \citenamefont {T\"orm\"a}}]{Visuri2014a}%
  \BibitemOpen
  \bibfield  {author} {\bibinfo {author} {\bibfnamefont {A.-M.}\ \bibnamefont
  {Visuri}}, \bibinfo {author} {\bibfnamefont {D.-H.}\ \bibnamefont {Kim}},
  \bibinfo {author} {\bibfnamefont {J.~J.}\ \bibnamefont {Kinnunen}}, \bibinfo
  {author} {\bibfnamefont {F.}~\bibnamefont {Massel}}, \ and\ \bibinfo {author}
  {\bibfnamefont {P.}~\bibnamefont {T\"orm\"a}},\ }\href {\doibase
  10.1103/PhysRevA.90.051603} {\bibfield  {journal} {\bibinfo  {journal} {Phys.
  Rev. A}\ }\textbf {\bibinfo {volume} {90}},\ \bibinfo {pages} {051603}
  (\bibinfo {year} {2014})}\BibitemShut {NoStop}%
\bibitem [{\citenamefont {Nascimb\`ene}\ \emph {et~al.}(2009)\citenamefont
  {Nascimb\`ene}, \citenamefont {Navon}, \citenamefont {Jiang}, \citenamefont
  {Tarruell}, \citenamefont {Teichmann}, \citenamefont {McKeever},
  \citenamefont {Chevy},\ and\ \citenamefont {Salomon}}]{Nascimbene2009a}%
  \BibitemOpen
  \bibfield  {author} {\bibinfo {author} {\bibfnamefont {S.}~\bibnamefont
  {Nascimb\`ene}}, \bibinfo {author} {\bibfnamefont {N.}~\bibnamefont {Navon}},
  \bibinfo {author} {\bibfnamefont {K.~J.}\ \bibnamefont {Jiang}}, \bibinfo
  {author} {\bibfnamefont {L.}~\bibnamefont {Tarruell}}, \bibinfo {author}
  {\bibfnamefont {M.}~\bibnamefont {Teichmann}}, \bibinfo {author}
  {\bibfnamefont {J.}~\bibnamefont {McKeever}}, \bibinfo {author}
  {\bibfnamefont {F.}~\bibnamefont {Chevy}}, \ and\ \bibinfo {author}
  {\bibfnamefont {C.}~\bibnamefont {Salomon}},\ }\href {\doibase
  10.1103/PhysRevLett.103.170402} {\bibfield  {journal} {\bibinfo  {journal}
  {Phys. Rev. Lett.}\ }\textbf {\bibinfo {volume} {103}},\ \bibinfo {pages}
  {170402} (\bibinfo {year} {2009})}\BibitemShut {NoStop}%
\bibitem [{\citenamefont {Schirotzek}\ \emph {et~al.}(2009)\citenamefont
  {Schirotzek}, \citenamefont {Wu}, \citenamefont {Sommer},\ and\ \citenamefont
  {Zwierlein}}]{Schirotzek2009a}%
  \BibitemOpen
  \bibfield  {author} {\bibinfo {author} {\bibfnamefont {A.}~\bibnamefont
  {Schirotzek}}, \bibinfo {author} {\bibfnamefont {C.-H.}\ \bibnamefont {Wu}},
  \bibinfo {author} {\bibfnamefont {A.}~\bibnamefont {Sommer}}, \ and\ \bibinfo
  {author} {\bibfnamefont {M.~W.}\ \bibnamefont {Zwierlein}},\ }\href {\doibase
  10.1103/PhysRevLett.102.230402} {\bibfield  {journal} {\bibinfo  {journal}
  {Phys. Rev. Lett.}\ }\textbf {\bibinfo {volume} {102}},\ \bibinfo {pages}
  {230402} (\bibinfo {year} {2009})}\BibitemShut {NoStop}%
\bibitem [{\citenamefont {Chevy}(2006)}]{Chevy2006a}%
  \BibitemOpen
  \bibfield  {author} {\bibinfo {author} {\bibfnamefont {F.}~\bibnamefont
  {Chevy}},\ }\href {\doibase 10.1103/PhysRevA.74.063628} {\bibfield  {journal}
  {\bibinfo  {journal} {Phys. Rev. A}\ }\textbf {\bibinfo {volume} {74}},\
  \bibinfo {pages} {063628} (\bibinfo {year} {2006})}\BibitemShut {NoStop}%
\bibitem [{\citenamefont {Giraud}\ and\ \citenamefont
  {Combescot}(2009)}]{Giraud2009a}%
  \BibitemOpen
  \bibfield  {author} {\bibinfo {author} {\bibfnamefont {S.}~\bibnamefont
  {Giraud}}\ and\ \bibinfo {author} {\bibfnamefont {R.}~\bibnamefont
  {Combescot}},\ }\href {\doibase 10.1103/PhysRevA.79.043615} {\bibfield
  {journal} {\bibinfo  {journal} {Phys. Rev. A.}\ }\textbf {\bibinfo {volume}
  {79}},\ \bibinfo {pages} {043615} (\bibinfo {year} {2009})}\BibitemShut
  {NoStop}%
\bibitem [{\citenamefont {Doggen}\ and\ \citenamefont
  {Kinnunen}(2013)}]{Doggen2013a}%
  \BibitemOpen
  \bibfield  {author} {\bibinfo {author} {\bibfnamefont {E.~V.~H.}\
  \bibnamefont {Doggen}}\ and\ \bibinfo {author} {\bibfnamefont {J.~J.}\
  \bibnamefont {Kinnunen}},\ }\href {\doibase 10.1103/PhysRevLett.111.025302}
  {\bibfield  {journal} {\bibinfo  {journal} {Phys. Rev. Lett.}\ }\textbf
  {\bibinfo {volume} {111}},\ \bibinfo {pages} {025302} (\bibinfo {year}
  {2013})}\BibitemShut {NoStop}%
\bibitem [{\citenamefont {Doggen}\ \emph {et~al.}(2014)\citenamefont {Doggen},
  \citenamefont {Korolyuk}, \citenamefont {T\"orm\"a},\ and\ \citenamefont
  {Kinnunen}}]{Doggen2014a}%
  \BibitemOpen
  \bibfield  {author} {\bibinfo {author} {\bibfnamefont {E.~V.~H.}\
  \bibnamefont {Doggen}}, \bibinfo {author} {\bibfnamefont {A.}~\bibnamefont
  {Korolyuk}}, \bibinfo {author} {\bibfnamefont {P.}~\bibnamefont {T\"orm\"a}},
  \ and\ \bibinfo {author} {\bibfnamefont {J.~J.}\ \bibnamefont {Kinnunen}},\
  }\href {\doibase 10.1103/PhysRevA.89.053621} {\bibfield  {journal} {\bibinfo
  {journal} {Phys. Rev. A}\ }\textbf {\bibinfo {volume} {89}},\ \bibinfo
  {pages} {053621} (\bibinfo {year} {2014})}\BibitemShut {NoStop}%
\bibitem [{\citenamefont {Rath}\ and\ \citenamefont
  {Schmidt}(2013)}]{Rath2013a}%
  \BibitemOpen
  \bibfield  {author} {\bibinfo {author} {\bibfnamefont {S.~P.}\ \bibnamefont
  {Rath}}\ and\ \bibinfo {author} {\bibfnamefont {R.}~\bibnamefont {Schmidt}},\
  }\href {\doibase 10.1103/PhysRevA.88.053632} {\bibfield  {journal} {\bibinfo
  {journal} {Phys. Rev. A}\ }\textbf {\bibinfo {volume} {88}},\ \bibinfo
  {pages} {053632} (\bibinfo {year} {2013})}\BibitemShut {NoStop}%
\bibitem [{\citenamefont {Dutta}\ and\ \citenamefont
  {Mueller}(2013)}]{Dutta2013a}%
  \BibitemOpen
  \bibfield  {author} {\bibinfo {author} {\bibfnamefont {S.}~\bibnamefont
  {Dutta}}\ and\ \bibinfo {author} {\bibfnamefont {E.~J.}\ \bibnamefont
  {Mueller}},\ }\href {\doibase 10.1103/PhysRevA.88.053601} {\bibfield
  {journal} {\bibinfo  {journal} {Phys. Rev. A}\ }\textbf {\bibinfo {volume}
  {88}},\ \bibinfo {pages} {053601} (\bibinfo {year} {2013})}\BibitemShut
  {NoStop}%
\bibitem [{\citenamefont {Li}\ and\ \citenamefont {Das~Sarma}(2014)}]{Li2014a}%
  \BibitemOpen
  \bibfield  {author} {\bibinfo {author} {\bibfnamefont {W.}~\bibnamefont
  {Li}}\ and\ \bibinfo {author} {\bibfnamefont {S.}~\bibnamefont {Das~Sarma}},\
  }\href {\doibase 10.1103/PhysRevA.90.013618} {\bibfield  {journal} {\bibinfo
  {journal} {Phys. Rev. A}\ }\textbf {\bibinfo {volume} {90}},\ \bibinfo
  {pages} {013618} (\bibinfo {year} {2014})}\BibitemShut {NoStop}%
\bibitem [{\citenamefont {Li}\ \emph {et~al.}(2003)\citenamefont {Li},
  \citenamefont {Gu}, \citenamefont {Ying},\ and\ \citenamefont
  {Eckern}}]{Li2003a}%
  \BibitemOpen
  \bibfield  {author} {\bibinfo {author} {\bibfnamefont {Y.-Q.}\ \bibnamefont
  {Li}}, \bibinfo {author} {\bibfnamefont {S.-J.}\ \bibnamefont {Gu}}, \bibinfo
  {author} {\bibfnamefont {Z.-J.}\ \bibnamefont {Ying}}, \ and\ \bibinfo
  {author} {\bibfnamefont {U.}~\bibnamefont {Eckern}},\ }\href
  {http://stacks.iop.org/0295-5075/61/i=3/a=368} {\bibfield  {journal}
  {\bibinfo  {journal} {Europhys. Lett.}\ }\textbf {\bibinfo {volume} {61}},\
  \bibinfo {pages} {368} (\bibinfo {year} {2003})}\BibitemShut {NoStop}%
\bibitem [{\citenamefont {Fuchs}\ \emph {et~al.}(2005)\citenamefont {Fuchs},
  \citenamefont {Gangardt}, \citenamefont {Keilmann},\ and\ \citenamefont
  {Shlyapnikov}}]{Fuchs2005a}%
  \BibitemOpen
  \bibfield  {author} {\bibinfo {author} {\bibfnamefont {J.~N.}\ \bibnamefont
  {Fuchs}}, \bibinfo {author} {\bibfnamefont {D.~M.}\ \bibnamefont {Gangardt}},
  \bibinfo {author} {\bibfnamefont {T.}~\bibnamefont {Keilmann}}, \ and\
  \bibinfo {author} {\bibfnamefont {G.~V.}\ \bibnamefont {Shlyapnikov}},\
  }\href {\doibase 10.1103/PhysRevLett.95.150402} {\bibfield  {journal}
  {\bibinfo  {journal} {Phys. Rev. Lett.}\ }\textbf {\bibinfo {volume} {95}},\
  \bibinfo {pages} {150402} (\bibinfo {year} {2005})}\BibitemShut {NoStop}%
\bibitem [{\citenamefont {Guan}\ \emph {et~al.}(2007)\citenamefont {Guan},
  \citenamefont {Batchelor},\ and\ \citenamefont {Takahashi}}]{Guan2007a}%
  \BibitemOpen
  \bibfield  {author} {\bibinfo {author} {\bibfnamefont {X.-W.}\ \bibnamefont
  {Guan}}, \bibinfo {author} {\bibfnamefont {M.~T.}\ \bibnamefont {Batchelor}},
  \ and\ \bibinfo {author} {\bibfnamefont {M.}~\bibnamefont {Takahashi}},\
  }\href {\doibase 10.1103/PhysRevA.76.043617} {\bibfield  {journal} {\bibinfo
  {journal} {Phys. Rev. A}\ }\textbf {\bibinfo {volume} {76}},\ \bibinfo
  {pages} {043617} (\bibinfo {year} {2007})}\BibitemShut {NoStop}%
\bibitem [{\citenamefont {Tempfli}\ \emph {et~al.}(2008)\citenamefont
  {Tempfli}, \citenamefont {Z\"ollner},\ and\ \citenamefont
  {Schmelcher}}]{Tempfli2008a}%
  \BibitemOpen
  \bibfield  {author} {\bibinfo {author} {\bibfnamefont {E.}~\bibnamefont
  {Tempfli}}, \bibinfo {author} {\bibfnamefont {S.}~\bibnamefont {Z\"ollner}},
  \ and\ \bibinfo {author} {\bibfnamefont {P.}~\bibnamefont {Schmelcher}},\
  }\href {http://stacks.iop.org/1367-2630/10/i=10/a=103021} {\bibfield
  {journal} {\bibinfo  {journal} {New J. Phys.}\ }\textbf {\bibinfo {volume}
  {10}},\ \bibinfo {pages} {103021} (\bibinfo {year} {2008})}\BibitemShut
  {NoStop}%
\bibitem [{\citenamefont {Tempfli}\ \emph {et~al.}(2009)\citenamefont
  {Tempfli}, \citenamefont {Z\"ollner},\ and\ \citenamefont
  {Schmelcher}}]{Tempfli2009a}%
  \BibitemOpen
  \bibfield  {author} {\bibinfo {author} {\bibfnamefont {E.}~\bibnamefont
  {Tempfli}}, \bibinfo {author} {\bibfnamefont {S.}~\bibnamefont {Z\"ollner}},
  \ and\ \bibinfo {author} {\bibfnamefont {P.}~\bibnamefont {Schmelcher}},\
  }\href {http://stacks.iop.org/1367-2630/11/i=7/a=073015} {\bibfield
  {journal} {\bibinfo  {journal} {New J. Phys.}\ }\textbf {\bibinfo {volume}
  {11}},\ \bibinfo {pages} {073015} (\bibinfo {year} {2009})}\BibitemShut
  {NoStop}%
\bibitem [{\citenamefont {Caux}\ \emph {et~al.}(2009)\citenamefont {Caux},
  \citenamefont {Klauser},\ and\ \citenamefont {van~den Brink}}]{Caux2009a}%
  \BibitemOpen
  \bibfield  {author} {\bibinfo {author} {\bibfnamefont {J.-S.}\ \bibnamefont
  {Caux}}, \bibinfo {author} {\bibfnamefont {A.}~\bibnamefont {Klauser}}, \
  and\ \bibinfo {author} {\bibfnamefont {J.}~\bibnamefont {van~den Brink}},\
  }\href {\doibase 10.1103/PhysRevA.80.061605} {\bibfield  {journal} {\bibinfo
  {journal} {Phys. Rev. A}\ }\textbf {\bibinfo {volume} {80}},\ \bibinfo
  {pages} {061605} (\bibinfo {year} {2009})}\BibitemShut {NoStop}%
\bibitem [{\citenamefont {Brouzos}\ and\ \citenamefont
  {Schmelcher}(2012)}]{Brouzos2012a}%
  \BibitemOpen
  \bibfield  {author} {\bibinfo {author} {\bibfnamefont {I.}~\bibnamefont
  {Brouzos}}\ and\ \bibinfo {author} {\bibfnamefont {P.}~\bibnamefont
  {Schmelcher}},\ }\href {\doibase 10.1103/PhysRevLett.108.045301} {\bibfield
  {journal} {\bibinfo  {journal} {Phys. Rev. Lett.}\ }\textbf {\bibinfo
  {volume} {108}},\ \bibinfo {pages} {045301} (\bibinfo {year}
  {2012})}\BibitemShut {NoStop}%
\bibitem [{\citenamefont {Vidmar}\ \emph {et~al.}(2013)\citenamefont {Vidmar},
  \citenamefont {Langer}, \citenamefont {McCulloch}, \citenamefont {Schneider},
  \citenamefont {Schollw\"ock},\ and\ \citenamefont
  {Heidrich-Meisner}}]{Vidmar2013a}%
  \BibitemOpen
  \bibfield  {author} {\bibinfo {author} {\bibfnamefont {L.}~\bibnamefont
  {Vidmar}}, \bibinfo {author} {\bibfnamefont {S.}~\bibnamefont {Langer}},
  \bibinfo {author} {\bibfnamefont {I.~P.}\ \bibnamefont {McCulloch}}, \bibinfo
  {author} {\bibfnamefont {U.}~\bibnamefont {Schneider}}, \bibinfo {author}
  {\bibfnamefont {U.}~\bibnamefont {Schollw\"ock}}, \ and\ \bibinfo {author}
  {\bibfnamefont {F.}~\bibnamefont {Heidrich-Meisner}},\ }\href {\doibase
  10.1103/PhysRevB.88.235117} {\bibfield  {journal} {\bibinfo  {journal} {Phys.
  Rev. B}\ }\textbf {\bibinfo {volume} {88}},\ \bibinfo {pages} {235117}
  (\bibinfo {year} {2013})}\BibitemShut {NoStop}%
\bibitem [{\citenamefont {Campbell}\ \emph {et~al.}(2014)\citenamefont
  {Campbell}, \citenamefont {Garc\'ia-March}, \citenamefont {Fogarty},\ and\
  \citenamefont {Busch}}]{Campbell2014a}%
  \BibitemOpen
  \bibfield  {author} {\bibinfo {author} {\bibfnamefont {S.}~\bibnamefont
  {Campbell}}, \bibinfo {author} {\bibfnamefont {M.~A.}\ \bibnamefont
  {Garc\'ia-March}}, \bibinfo {author} {\bibfnamefont {T.}~\bibnamefont
  {Fogarty}}, \ and\ \bibinfo {author} {\bibfnamefont {T.}~\bibnamefont
  {Busch}},\ }\href {\doibase 10.1103/PhysRevA.90.013617} {\bibfield  {journal}
  {\bibinfo  {journal} {Phys. Rev. A}\ }\textbf {\bibinfo {volume} {90}},\
  \bibinfo {pages} {013617} (\bibinfo {year} {2014})}\BibitemShut {NoStop}%
\bibitem [{\citenamefont {Boschi}\ \emph {et~al.}(2014)\citenamefont {Boschi},
  \citenamefont {Ercolessi}, \citenamefont {Ferrari}, \citenamefont {Naldesi},
  \citenamefont {Ortolani},\ and\ \citenamefont {Taddia}}]{Boschi2014a}%
  \BibitemOpen
  \bibfield  {author} {\bibinfo {author} {\bibfnamefont {C.~D.~E.}\
  \bibnamefont {Boschi}}, \bibinfo {author} {\bibfnamefont {E.}~\bibnamefont
  {Ercolessi}}, \bibinfo {author} {\bibfnamefont {L.}~\bibnamefont {Ferrari}},
  \bibinfo {author} {\bibfnamefont {P.}~\bibnamefont {Naldesi}}, \bibinfo
  {author} {\bibfnamefont {F.}~\bibnamefont {Ortolani}}, \ and\ \bibinfo
  {author} {\bibfnamefont {L.}~\bibnamefont {Taddia}},\ }\href {\doibase
  10.1103/PhysRevA.90.043606} {\bibfield  {journal} {\bibinfo  {journal} {Phys.
  Rev. A}\ }\textbf {\bibinfo {volume} {90}},\ \bibinfo {pages} {043606}
  (\bibinfo {year} {2014})}\BibitemShut {NoStop}%
\bibitem [{\citenamefont {Zinner}\ \emph {et~al.}(2014)\citenamefont {Zinner},
  \citenamefont {Volosniev}, \citenamefont {Fedorov}, \citenamefont {Jensen},\
  and\ \citenamefont {Valiente}}]{Zinner2014a}%
  \BibitemOpen
  \bibfield  {author} {\bibinfo {author} {\bibfnamefont {N.~T.}\ \bibnamefont
  {Zinner}}, \bibinfo {author} {\bibfnamefont {A.~G.}\ \bibnamefont
  {Volosniev}}, \bibinfo {author} {\bibfnamefont {D.~V.}\ \bibnamefont
  {Fedorov}}, \bibinfo {author} {\bibfnamefont {A.~S.}\ \bibnamefont {Jensen}},
  \ and\ \bibinfo {author} {\bibfnamefont {M.}~\bibnamefont {Valiente}},\
  }\href {http://stacks.iop.org/0295-5075/107/i=6/a=60003} {\bibfield
  {journal} {\bibinfo  {journal} {Europhys. Lett.}\ }\textbf {\bibinfo {volume}
  {107}},\ \bibinfo {pages} {60003} (\bibinfo {year} {2014})}\BibitemShut
  {NoStop}%
\bibitem [{\citenamefont {Campbell}\ \emph {et~al.}(2015)\citenamefont
  {Campbell}, \citenamefont {Gangardt},\ and\ \citenamefont
  {Kheruntsyan}}]{Campbell2015a}%
  \BibitemOpen
  \bibfield  {author} {\bibinfo {author} {\bibfnamefont {A.~S.}\ \bibnamefont
  {Campbell}}, \bibinfo {author} {\bibfnamefont {D.~M.}\ \bibnamefont
  {Gangardt}}, \ and\ \bibinfo {author} {\bibfnamefont {K.~V.}\ \bibnamefont
  {Kheruntsyan}},\ }\href {\doibase 10.1103/PhysRevLett.114.125302} {\bibfield
  {journal} {\bibinfo  {journal} {Phys. Rev. Lett.}\ }\textbf {\bibinfo
  {volume} {114}},\ \bibinfo {pages} {125302} (\bibinfo {year}
  {2015})}\BibitemShut {NoStop}%
\bibitem [{\citenamefont {Rigol}\ \emph {et~al.}(2007)\citenamefont {Rigol},
  \citenamefont {Dunjko}, \citenamefont {Yurovsky},\ and\ \citenamefont
  {Olshanii}}]{Rigol2007a}%
  \BibitemOpen
  \bibfield  {author} {\bibinfo {author} {\bibfnamefont {M.}~\bibnamefont
  {Rigol}}, \bibinfo {author} {\bibfnamefont {V.}~\bibnamefont {Dunjko}},
  \bibinfo {author} {\bibfnamefont {V.}~\bibnamefont {Yurovsky}}, \ and\
  \bibinfo {author} {\bibfnamefont {M.}~\bibnamefont {Olshanii}},\ }\href
  {\doibase 10.1103/PhysRevLett.98.050405} {\bibfield  {journal} {\bibinfo
  {journal} {Phys. Rev. Lett.}\ }\textbf {\bibinfo {volume} {98}},\ \bibinfo
  {pages} {050405} (\bibinfo {year} {2007})}\BibitemShut {NoStop}%
\bibitem [{\citenamefont {Rigol}\ \emph {et~al.}(2008)\citenamefont {Rigol},
  \citenamefont {Dunjko},\ and\ \citenamefont {Olshanii}}]{Rigol2008a}%
  \BibitemOpen
  \bibfield  {author} {\bibinfo {author} {\bibfnamefont {M.}~\bibnamefont
  {Rigol}}, \bibinfo {author} {\bibfnamefont {V.}~\bibnamefont {Dunjko}}, \
  and\ \bibinfo {author} {\bibfnamefont {M.}~\bibnamefont {Olshanii}},\ }\href
  {\doibase doi:10.1038/nature06838} {\bibfield  {journal} {\bibinfo  {journal}
  {Nature}\ }\textbf {\bibinfo {volume} {452}},\ \bibinfo {pages} {854}
  (\bibinfo {year} {2008})}\BibitemShut {NoStop}%
\bibitem [{\citenamefont {Cazalilla}\ and\ \citenamefont
  {Rigol}(2010)}]{Cazalilla2010b}%
  \BibitemOpen
  \bibfield  {author} {\bibinfo {author} {\bibfnamefont {M.~A.}\ \bibnamefont
  {Cazalilla}}\ and\ \bibinfo {author} {\bibfnamefont {M.}~\bibnamefont
  {Rigol}},\ }\href {http://stacks.iop.org/1367-2630/12/i=5/a=055006}
  {\bibfield  {journal} {\bibinfo  {journal} {New J. Phys.}\ }\textbf {\bibinfo
  {volume} {12}},\ \bibinfo {pages} {055006} (\bibinfo {year}
  {2010})}\BibitemShut {NoStop}%
\bibitem [{\citenamefont {Polkovnikov}\ \emph {et~al.}(2011)\citenamefont
  {Polkovnikov}, \citenamefont {Sengupta}, \citenamefont {Silva},\ and\
  \citenamefont {Vengalattore}}]{Polkovnikov2011a}%
  \BibitemOpen
  \bibfield  {author} {\bibinfo {author} {\bibfnamefont {A.}~\bibnamefont
  {Polkovnikov}}, \bibinfo {author} {\bibfnamefont {K.}~\bibnamefont
  {Sengupta}}, \bibinfo {author} {\bibfnamefont {A.}~\bibnamefont {Silva}}, \
  and\ \bibinfo {author} {\bibfnamefont {M.}~\bibnamefont {Vengalattore}},\
  }\href {\doibase 10.1103/RevModPhys.83.863} {\bibfield  {journal} {\bibinfo
  {journal} {Rev. Mod. Phys.}\ }\textbf {\bibinfo {volume} {83}},\ \bibinfo
  {pages} {863} (\bibinfo {year} {2011})}\BibitemShut {NoStop}%
\bibitem [{\citenamefont {Meyrath}\ \emph {et~al.}(2005)\citenamefont
  {Meyrath}, \citenamefont {Schreck}, \citenamefont {Hanssen}, \citenamefont
  {Chuu},\ and\ \citenamefont {Raizen}}]{Meyrath2005a}%
  \BibitemOpen
  \bibfield  {author} {\bibinfo {author} {\bibfnamefont {T.~P.}\ \bibnamefont
  {Meyrath}}, \bibinfo {author} {\bibfnamefont {F.}~\bibnamefont {Schreck}},
  \bibinfo {author} {\bibfnamefont {J.~L.}\ \bibnamefont {Hanssen}}, \bibinfo
  {author} {\bibfnamefont {C.-S.}\ \bibnamefont {Chuu}}, \ and\ \bibinfo
  {author} {\bibfnamefont {M.~G.}\ \bibnamefont {Raizen}},\ }\href {\doibase
  10.1103/PhysRevA.71.041604} {\bibfield  {journal} {\bibinfo  {journal} {Phys.
  Rev. A}\ }\textbf {\bibinfo {volume} {71}},\ \bibinfo {pages} {041604}
  (\bibinfo {year} {2005})}\BibitemShut {NoStop}%
\bibitem [{\citenamefont {Z\"ollner}\ \emph {et~al.}(2008)\citenamefont
  {Z\"ollner}, \citenamefont {Meyer},\ and\ \citenamefont
  {Schmelcher}}]{Zollner2008a}%
  \BibitemOpen
  \bibfield  {author} {\bibinfo {author} {\bibfnamefont {S.}~\bibnamefont
  {Z\"ollner}}, \bibinfo {author} {\bibfnamefont {H.-D.}\ \bibnamefont
  {Meyer}}, \ and\ \bibinfo {author} {\bibfnamefont {P.}~\bibnamefont
  {Schmelcher}},\ }\href {\doibase 10.1103/PhysRevLett.100.040401} {\bibfield
  {journal} {\bibinfo  {journal} {Phys. Rev. Lett.}\ }\textbf {\bibinfo
  {volume} {100}},\ \bibinfo {pages} {040401} (\bibinfo {year}
  {2008})}\BibitemShut {NoStop}%
\bibitem [{\citenamefont {Rontani}(2012)}]{Rontani2012a}%
  \BibitemOpen
  \bibfield  {author} {\bibinfo {author} {\bibfnamefont {M.}~\bibnamefont
  {Rontani}},\ }\href {\doibase 10.1103/PhysRevLett.108.115302} {\bibfield
  {journal} {\bibinfo  {journal} {Phys. Rev. Lett.}\ }\textbf {\bibinfo
  {volume} {108}},\ \bibinfo {pages} {115302} (\bibinfo {year}
  {2012})}\BibitemShut {NoStop}%
\bibitem [{\citenamefont {Simpson}\ \emph {et~al.}(2014)\citenamefont
  {Simpson}, \citenamefont {Gangardt}, \citenamefont {Lerner},\ and\
  \citenamefont {Kr\"uger}}]{Simpson2014a}%
  \BibitemOpen
  \bibfield  {author} {\bibinfo {author} {\bibfnamefont {D.~P.}\ \bibnamefont
  {Simpson}}, \bibinfo {author} {\bibfnamefont {D.~M.}\ \bibnamefont
  {Gangardt}}, \bibinfo {author} {\bibfnamefont {I.~V.}\ \bibnamefont
  {Lerner}}, \ and\ \bibinfo {author} {\bibfnamefont {P.}~\bibnamefont
  {Kr\"uger}},\ }\href {\doibase 10.1103/PhysRevLett.112.100601} {\bibfield
  {journal} {\bibinfo  {journal} {Phys. Rev. Lett.}\ }\textbf {\bibinfo
  {volume} {112}},\ \bibinfo {pages} {100601} (\bibinfo {year}
  {2014})}\BibitemShut {NoStop}%
\bibitem [{\citenamefont {Lundmark}\ \emph {et~al.}(2015)\citenamefont
  {Lundmark}, \citenamefont {Forss\'en},\ and\ \citenamefont
  {Rotureau}}]{Lundmark2015a}%
  \BibitemOpen
  \bibfield  {author} {\bibinfo {author} {\bibfnamefont {R.}~\bibnamefont
  {Lundmark}}, \bibinfo {author} {\bibfnamefont {C.}~\bibnamefont {Forss\'en}},
  \ and\ \bibinfo {author} {\bibfnamefont {J.}~\bibnamefont {Rotureau}},\
  }\href {\doibase 10.1103/PhysRevA.91.041601} {\bibfield  {journal} {\bibinfo
  {journal} {Phys. Rev. A}\ }\textbf {\bibinfo {volume} {91}},\ \bibinfo
  {pages} {041601} (\bibinfo {year} {2015})}\BibitemShut {NoStop}%
\bibitem [{\citenamefont {Sommer}\ \emph {et~al.}(2011)\citenamefont {Sommer},
  \citenamefont {Ku}, \citenamefont {Roati},\ and\ \citenamefont
  {Zwierlein}}]{Sommer2011a}%
  \BibitemOpen
  \bibfield  {author} {\bibinfo {author} {\bibfnamefont {A.}~\bibnamefont
  {Sommer}}, \bibinfo {author} {\bibfnamefont {M.}~\bibnamefont {Ku}}, \bibinfo
  {author} {\bibfnamefont {G.}~\bibnamefont {Roati}}, \ and\ \bibinfo {author}
  {\bibfnamefont {M.~W.}\ \bibnamefont {Zwierlein}},\ }\href@noop {} {\bibfield
   {journal} {\bibinfo  {journal} {Nature}\ }\textbf {\bibinfo {volume}
  {472}},\ \bibinfo {pages} {201} (\bibinfo {year} {2011})}\BibitemShut
  {NoStop}%
\bibitem [{\citenamefont {Will}\ \emph {et~al.}(2011)\citenamefont {Will},
  \citenamefont {Best}, \citenamefont {Braun}, \citenamefont {Schneider},\ and\
  \citenamefont {Bloch}}]{Will2011a}%
  \BibitemOpen
  \bibfield  {author} {\bibinfo {author} {\bibfnamefont {S.}~\bibnamefont
  {Will}}, \bibinfo {author} {\bibfnamefont {T.}~\bibnamefont {Best}}, \bibinfo
  {author} {\bibfnamefont {S.}~\bibnamefont {Braun}}, \bibinfo {author}
  {\bibfnamefont {U.}~\bibnamefont {Schneider}}, \ and\ \bibinfo {author}
  {\bibfnamefont {I.}~\bibnamefont {Bloch}},\ }\href {\doibase
  10.1103/PhysRevLett.106.115305} {\bibfield  {journal} {\bibinfo  {journal}
  {Phys. Rev. Lett.}\ }\textbf {\bibinfo {volume} {106}},\ \bibinfo {pages}
  {115305} (\bibinfo {year} {2011})}\BibitemShut {NoStop}%
\bibitem [{\citenamefont {Lieb}\ and\ \citenamefont
  {Liniger}(1963)}]{Lieb1963a}%
  \BibitemOpen
  \bibfield  {author} {\bibinfo {author} {\bibfnamefont {E.~H.}\ \bibnamefont
  {Lieb}}\ and\ \bibinfo {author} {\bibfnamefont {W.}~\bibnamefont {Liniger}},\
  }\href {\doibase 10.1103/PhysRev.130.1605} {\bibfield  {journal} {\bibinfo
  {journal} {Phys. Rev.}\ }\textbf {\bibinfo {volume} {130}},\ \bibinfo {pages}
  {1605} (\bibinfo {year} {1963})}\BibitemShut {NoStop}%
\bibitem [{\citenamefont {LeBlanc}\ and\ \citenamefont
  {Thywissen}(2007)}]{LeBlanc2007a}%
  \BibitemOpen
  \bibfield  {author} {\bibinfo {author} {\bibfnamefont {L.~J.}\ \bibnamefont
  {LeBlanc}}\ and\ \bibinfo {author} {\bibfnamefont {J.~H.}\ \bibnamefont
  {Thywissen}},\ }\href {\doibase 10.1103/PhysRevA.75.053612} {\bibfield
  {journal} {\bibinfo  {journal} {Phys. Rev. A}\ }\textbf {\bibinfo {volume}
  {75}},\ \bibinfo {pages} {053612} (\bibinfo {year} {2007})}\BibitemShut
  {NoStop}%
\bibitem [{\citenamefont {Goulko}\ \emph {et~al.}(2011)\citenamefont {Goulko},
  \citenamefont {Chevy},\ and\ \citenamefont {Lobo}}]{Goulko2011a}%
  \BibitemOpen
  \bibfield  {author} {\bibinfo {author} {\bibfnamefont {O.}~\bibnamefont
  {Goulko}}, \bibinfo {author} {\bibfnamefont {F.}~\bibnamefont {Chevy}}, \
  and\ \bibinfo {author} {\bibfnamefont {C.}~\bibnamefont {Lobo}},\ }\href
  {\doibase 10.1103/PhysRevA.84.051605} {\bibfield  {journal} {\bibinfo
  {journal} {Phys. Rev. A}\ }\textbf {\bibinfo {volume} {84}},\ \bibinfo
  {pages} {051605} (\bibinfo {year} {2011})}\BibitemShut {NoStop}%
\bibitem [{\citenamefont {Ozaki}\ \emph {et~al.}(2012)\citenamefont {Ozaki},
  \citenamefont {Tezuka},\ and\ \citenamefont {Kawakami}}]{Ozaki2012a}%
  \BibitemOpen
  \bibfield  {author} {\bibinfo {author} {\bibfnamefont {J.}~\bibnamefont
  {Ozaki}}, \bibinfo {author} {\bibfnamefont {M.}~\bibnamefont {Tezuka}}, \
  and\ \bibinfo {author} {\bibfnamefont {N.}~\bibnamefont {Kawakami}},\ }\href
  {\doibase 10.1103/PhysRevA.86.033621} {\bibfield  {journal} {\bibinfo
  {journal} {Phys. Rev. A}\ }\textbf {\bibinfo {volume} {86}},\ \bibinfo
  {pages} {033621} (\bibinfo {year} {2012})}\BibitemShut {NoStop}%
\bibitem [{\citenamefont {Doggen}\ and\ \citenamefont
  {Kinnunen}(2014)}]{Doggen2014b}%
  \BibitemOpen
  \bibfield  {author} {\bibinfo {author} {\bibfnamefont {E.~V.~H.}\
  \bibnamefont {Doggen}}\ and\ \bibinfo {author} {\bibfnamefont {J.~J.}\
  \bibnamefont {Kinnunen}},\ }\href@noop {} {\bibfield  {journal} {\bibinfo
  {journal} {New J. Phys.}\ }\textbf {\bibinfo {volume} {16}},\ \bibinfo
  {pages} {113051} (\bibinfo {year} {2014})}\BibitemShut {NoStop}%
\bibitem [{\citenamefont {Taylor}\ \emph {et~al.}(2011)\citenamefont {Taylor},
  \citenamefont {Zhang}, \citenamefont {Schneider},\ and\ \citenamefont
  {Randeria}}]{Taylor2011a}%
  \BibitemOpen
  \bibfield  {author} {\bibinfo {author} {\bibfnamefont {E.}~\bibnamefont
  {Taylor}}, \bibinfo {author} {\bibfnamefont {S.}~\bibnamefont {Zhang}},
  \bibinfo {author} {\bibfnamefont {W.}~\bibnamefont {Schneider}}, \ and\
  \bibinfo {author} {\bibfnamefont {M.}~\bibnamefont {Randeria}},\ }\href
  {\doibase 10.1103/PhysRevA.84.063622} {\bibfield  {journal} {\bibinfo
  {journal} {Phys. Rev. A}\ }\textbf {\bibinfo {volume} {84}},\ \bibinfo
  {pages} {063622} (\bibinfo {year} {2011})}\BibitemShut {NoStop}%
\bibitem [{\citenamefont {Kajala}\ \emph
  {et~al.}(2011{\natexlab{a}})\citenamefont {Kajala}, \citenamefont {Massel},\
  and\ \citenamefont {T\"orm\"a}}]{Kajala2011a}%
  \BibitemOpen
  \bibfield  {author} {\bibinfo {author} {\bibfnamefont {J.}~\bibnamefont
  {Kajala}}, \bibinfo {author} {\bibfnamefont {F.}~\bibnamefont {Massel}}, \
  and\ \bibinfo {author} {\bibfnamefont {P.}~\bibnamefont {T\"orm\"a}},\ }\href
  {\doibase 10.1103/PhysRevLett.106.206401} {\bibfield  {journal} {\bibinfo
  {journal} {Phys. Rev. Lett.}\ }\textbf {\bibinfo {volume} {106}},\ \bibinfo
  {pages} {206401} (\bibinfo {year} {2011}{\natexlab{a}})}\BibitemShut
  {NoStop}%
\bibitem [{\citenamefont {Kajala}\ \emph
  {et~al.}(2011{\natexlab{b}})\citenamefont {Kajala}, \citenamefont {Massel},\
  and\ \citenamefont {T\"orm\"a}}]{Kajala2011b}%
  \BibitemOpen
  \bibfield  {author} {\bibinfo {author} {\bibfnamefont {J.}~\bibnamefont
  {Kajala}}, \bibinfo {author} {\bibfnamefont {F.}~\bibnamefont {Massel}}, \
  and\ \bibinfo {author} {\bibfnamefont {P.}~\bibnamefont {T\"orm\"a}},\ }\href
  {\doibase 10.1140/epjd/e2011-20081-8} {\bibfield  {journal} {\bibinfo
  {journal} {Eur. Phys. J. D}\ }\textbf {\bibinfo {volume} {65}},\ \bibinfo
  {pages} {91} (\bibinfo {year} {2011}{\natexlab{b}})}\BibitemShut {NoStop}%
\bibitem [{\citenamefont {Schneider}\ \emph {et~al.}(2012)\citenamefont
  {Schneider}, \citenamefont {Hackerm\"uller}, \citenamefont {Ronzheimer},
  \citenamefont {Will}, \citenamefont {Braun}, \citenamefont {Best},
  \citenamefont {Bloch}, \citenamefont {Demler}, \citenamefont {Mandt},
  \citenamefont {Rasch},\ and\ \citenamefont {Rosch}}]{Schneider2012a}%
  \BibitemOpen
  \bibfield  {author} {\bibinfo {author} {\bibfnamefont {U.}~\bibnamefont
  {Schneider}}, \bibinfo {author} {\bibfnamefont {L.}~\bibnamefont
  {Hackerm\"uller}}, \bibinfo {author} {\bibfnamefont {J.~P.}\ \bibnamefont
  {Ronzheimer}}, \bibinfo {author} {\bibfnamefont {S.}~\bibnamefont {Will}},
  \bibinfo {author} {\bibfnamefont {S.}~\bibnamefont {Braun}}, \bibinfo
  {author} {\bibfnamefont {T.}~\bibnamefont {Best}}, \bibinfo {author}
  {\bibfnamefont {I.}~\bibnamefont {Bloch}}, \bibinfo {author} {\bibfnamefont
  {E.}~\bibnamefont {Demler}}, \bibinfo {author} {\bibfnamefont
  {S.}~\bibnamefont {Mandt}}, \bibinfo {author} {\bibfnamefont
  {D.}~\bibnamefont {Rasch}}, \ and\ \bibinfo {author} {\bibfnamefont
  {A.}~\bibnamefont {Rosch}},\ }\href {\doibase 10.1038/NPHYS2205} {\bibfield
  {journal} {\bibinfo  {journal} {Nature Phys.}\ }\textbf {\bibinfo {volume}
  {8}},\ \bibinfo {pages} {213} (\bibinfo {year} {2012})}\BibitemShut {NoStop}%
\bibitem [{\citenamefont {Rontani}(2013)}]{Rontani2013a}%
  \BibitemOpen
  \bibfield  {author} {\bibinfo {author} {\bibfnamefont {M.}~\bibnamefont
  {Rontani}},\ }\href {\doibase 10.1103/PhysRevA.88.043633} {\bibfield
  {journal} {\bibinfo  {journal} {Phys. Rev. A}\ }\textbf {\bibinfo {volume}
  {88}},\ \bibinfo {pages} {043633} (\bibinfo {year} {2013})}\BibitemShut
  {NoStop}%
\bibitem [{\citenamefont {Olshanii}\ and\ \citenamefont
  {Dunjko}(2003)}]{Olshanii2003a}%
  \BibitemOpen
  \bibfield  {author} {\bibinfo {author} {\bibfnamefont {M.}~\bibnamefont
  {Olshanii}}\ and\ \bibinfo {author} {\bibfnamefont {V.}~\bibnamefont
  {Dunjko}},\ }\href {\doibase 10.1103/PhysRevLett.91.090401} {\bibfield
  {journal} {\bibinfo  {journal} {Phys. Rev. Lett.}\ }\textbf {\bibinfo
  {volume} {91}},\ \bibinfo {pages} {090401} (\bibinfo {year}
  {2003})}\BibitemShut {NoStop}%
\bibitem [{\citenamefont {Anderson}\ \emph {et~al.}(1999)\citenamefont
  {Anderson}, \citenamefont {Bai}, \citenamefont {Bischof}, \citenamefont
  {Blackford}, \citenamefont {Demmel}, \citenamefont {Dongarra}, \citenamefont
  {Du~Croz}, \citenamefont {Greenbaum}, \citenamefont {Hammarling},
  \citenamefont {McKenney},\ and\ \citenamefont {Sorensen}}]{Lapack1999}%
  \BibitemOpen
  \bibfield  {author} {\bibinfo {author} {\bibfnamefont {E.}~\bibnamefont
  {Anderson}}, \bibinfo {author} {\bibfnamefont {Z.}~\bibnamefont {Bai}},
  \bibinfo {author} {\bibfnamefont {C.}~\bibnamefont {Bischof}}, \bibinfo
  {author} {\bibfnamefont {S.}~\bibnamefont {Blackford}}, \bibinfo {author}
  {\bibfnamefont {J.}~\bibnamefont {Demmel}}, \bibinfo {author} {\bibfnamefont
  {J.}~\bibnamefont {Dongarra}}, \bibinfo {author} {\bibfnamefont
  {J.}~\bibnamefont {Du~Croz}}, \bibinfo {author} {\bibfnamefont
  {A.}~\bibnamefont {Greenbaum}}, \bibinfo {author} {\bibfnamefont
  {S.}~\bibnamefont {Hammarling}}, \bibinfo {author} {\bibfnamefont
  {A.}~\bibnamefont {McKenney}}, \ and\ \bibinfo {author} {\bibfnamefont
  {D.}~\bibnamefont {Sorensen}},\ }\href@noop {} {\emph {\bibinfo {title}
  {{LAPACK} Users' Guide}}}\ (\bibinfo  {publisher} {Society for Industrial and
  Applied Mathematics},\ \bibinfo {address} {Philadelphia, PA},\ \bibinfo
  {year} {1999})\BibitemShut {NoStop}%
\end{thebibliography}%

\end{document}